\documentclass[reqno,a4paper]{article}
\usepackage{amssymb,amsmath,epsfig,float,amsfonts}
\usepackage{graphics,psfrag,graphicx,color}

\usepackage{subfigure,float,tabularx,ctable,fancyheadings,pstricks}

\definecolor{cgray}{rgb}{0.9,0.9,0.9}
\definecolor{lightblue}{rgb}{0.22,0.45,0.70}
\definecolor{lightred}{rgb}{0.8,0.25,0.2}

\usepackage[colorlinks=true,breaklinks=true,linkcolor=lightblue,citecolor=lightblue]{hyperref}
\numberwithin{equation}{section}

\newcommand{\mbs}[1]{\mathbf{#1}}

\def\beq{\begin{equation}} \def\eeq{\end{equation}}
\def\beqn{\begin{eqnarray}} \def\eeqn{\end{eqnarray}}

  \def\b0{{\mbs{0}}}

    \def\bu{{\mbs{u}}}

\graphicspath {{./}{./figures/}}

\allowdisplaybreaks
\title{A note on stress-driven anisotropic diffusion\\ and its role in active deformable media}
\author{\normalsize {\sc Christian Cherubini}\thanks{International Center for Relativistic Astrophysics, I.C.R.A. and 
Unit of Nonlinear Physics and Mathematical Modelling, Department of 
Engineering, University Campus Bio-Medico of Rome, Via A. del Portillo 21, 
00128 Rome, Italy. Email: {\tt \{c.cherubini,s.filippi\}@unicampus.it}.}\ ,\,  
{\sc Simonetta Filippi}$^*$,\,   
{\sc Alessio Gizzi}\thanks{Unit of Nonlinear Physics and Mathematical Modelling, Department of 
Engineering, University Campus Bio-Medico of Rome, Via A. del Portillo 21, 
00128 Rome, Italy.  Email: {\tt a.gizzi@unicampus.it}.}, \,  
{\sc Ricardo Ruiz Baier}\thanks{Mathematical Institute, University of Oxford,
A. Wiles Building, Woodstock Road, OX2 6GG Oxford, UK. Email: {\tt ruizbaier@maths.ox.ac.uk}.}
}
\date{December 20, 2016}
\begin{document}
\maketitle
% ****************************************************************************************
\begin{abstract}
We propose a new model to describe diffusion processes within active deformable media. Our general theoretical framework is based on physical and mathematical considerations, and it suggests to use diffusion tensors directly coupled to mechanical stress. {A proof-of-concept experiment and the proposed generalised reaction-diffusion-mechanics model} reveal that initially isotropic and homogeneous diffusion tensors turn into inhomogeneous and anisotropic {quantities} due to the intrinsic structure of the nonlinear coupling. We study the physical properties leading to these effects, and investigate mathematical conditions for its occurrence. Together, the experiment, the model, and the numerical results obtained using a mixed-primal finite element method, clearly support relevant consequences of stress-assisted diffusion into anisotropy patterns, drifting, and conduction velocity of the resulting excitation waves. Our findings also indicate the applicability of this novel approach in the description of mechano-electrical feedback in actively deforming bio-materials such as the heart.
\end{abstract}

\vspace{5pt}


%************************************************************************************************
\section{Introduction}
%************************************************************************************************
Excitable media, whether of biological type or not, represent complex
nonlinear systems. They are often of electrochemical nature, and  
can typically be coupled to  several 
multi-physical factors as heat transfer or solid and/or fluid
mechanics. A remarkable example is the heart, where 
nonlinear bioelectrical waves propagate on a complex anatomical
background undergoing large mechanical deformations and facing strong
interactions with biological fluids. More precisely,
cardiac contraction results from the combination of a complex emerging
behaviour where subcellular ion dynamics induce the overlapping of 
protein filaments, rapidly scaling up 
to both the cellular and tissue scales through a process known as
excitation-contraction coupling and, as main topic of the present work,  
its reverse effect known as the
\emph{mechano-electric feedback} (MEF)~\cite{kaufmann:1967,kohl:2001}. 
Studying the spatiotemporal dynamics of excitation waves in the heart 
is of paramount importance {in the understanding of a large class of 
processes including} depolarisation, repolarisation and period
doubling bifurcations occurring in the transition towards chaotic
regimes (arrhythmias) \cite{weiss:2009,das:2014,fenton:2017}.

Still in the context of cardiac dynamics, a large 
number of experimental data is available to describe ionic and electrophysiological 
processes at many spatio-temporal scales. 
However, due to diverse technical reasons, 
a common practice is to biochemically suppress any mechanical feedback to record 
these data, implying that any back-reacting effect 
intrinsically due to electromechanical interactions is systematically neglected. 
Nevertheless, the importance of understanding the interplay between the
reaction-diffusion (RD) dynamics with mechanical deformation is quite clear, and 
a recent growing interest 
in refining a companion mathematical model for the dynamics of higher
complexity models has been observed~\cite{quinn:2014a,quinn:2015,quinn:2014,ravelli:2003}. 
Even though several subcellular contributors to cardiac MEF have been 
extensively studied (as for instance, stretch-activated ion channels), 
their proper and consistent  integration into tissue-level models has remained a 
challenging task; 
further having a very limited clinically-translatable application and validation.

In the last fifteen years, the relation 
between cell or tissue-based electrophysiology models with
mechanical deformation of soft tissues has been formulated in terms of active
stress~\cite{nash:2004}, active 
strain~\cite{cherubini:2008,nobile:2012,gizzi:2015}, with the addition
of stretch-activated currents~\cite{panfilov:2005,panfilov:2007,quinn:2013,trayanova:2011}, or recently
combined with membrane capacitance changes~\cite{mcculloch:2015}.
However, in all of these works the key physical ingredient
ruling the spatio-temporal dynamics of the {membrane potential}, 
i.e. the diffusion tensor (in this context, the conductivity), 
has been systematically considered independent of mechanical 
deformation, {even if general constitutive prescriptions suggesting 
a possible interaction were advanced~\cite{nash:2004}}. 
Processes related to MEF have a fundamental role in a wide variety of 
\emph{passive} physical systems. Notable examples comprise
corrosion~\cite{zheng:2015}, rock anisotropy~\cite{johnson:1996},
glass transition~\cite{cohen:1989}, dissolution
phenomena~\cite{miller:2002}, 
electromigration~\cite{suo:1997}, hydrogen
trapping~\cite{chene:2008}, as well as 
swelling effects~\cite{suo:2009}. Clear evidence for the 
existence of such a coupling in biological systems has also been 
recently observed in strain-dependent 
oxygen diffusivity in cartilage~\cite{jackson:2009,yuan:2009}, and in
transcription factors within the cell nucleus~\cite{nava:2016}. 
Regarding the specific context of active biological media, 
connections forming gap junctions in cardiomyocytes (and associated 
to intercellular communication and mesoscale diffusion) have been 
recently discussed in terms of their mechano-sensitive properties~\cite{salameh:2013}. 
Furthermore, a quantitative analysis on the specific effects 
of stretch into connexins in terms of hemichannels has been experimentally 
verified in a number of different cellular preparations~\cite{bao:2004,cherian:2005}.

In the perspective of the present work, an important number of 
experimental studies have demonstrated key
MEF effects in ventricular myocardium and atrial tissue (see e.g.~\cite{ravelli:2003,quinn:2014} for an extended review). 
Specific applications include atria arrhythmias, where 
basically three main results are available. First, the spatiotemporal distribution of 
atrial excitation depends strongly on the anatomical substrate~\cite{ravelli:2005}. 
Secondly, upon mechanical loading (stretching),
the conduction velocity of the excitation wave decreases, a beat-to-beat 
interval variability appears, early afterdepolarisations, 
ectopic excitations and a higher vulnerability to atrial fibrillation are 
present~\cite{mase:2008}.
Third, atrial tissue undergoes multiple high frequency and unstable rotors 
(spiral waves)
when subject to constant and variable stretch states~\cite{jalife:2009}.
Experimental evidence of MEF has also been studied during ventricular 
loading, indicating a strong relationship between variations in the conduction 
velocity and strain anisotropy~\cite{lab:1978,franz:1989,mills:2011,mcculloch:2015}.

The fact that electrical properties of solids undergo intrinsic modifications 
due to (even infinitesimal) deformations has been a subject of study since 
a few decades (see e.g. the classical volume by  Landau
et al. on the electrodynamics of continuous media \cite[pp.~69]{landau:1984}). 
These changes suggest the representation of the corresponding models 
using strain-dependent (or also stress-dependent) dielectric 
tensors. In particular, this dependency in turn affects the electromagnetic dynamics
by enforcing inhomogeneous and anisotropic patterns in structures that  
were not necessarily so. 
Experimental evidence supports the main present assumption that  
 diffusion depends on deformation. Therefore, and thanks to first principles, 
 one can readily Taylor-expand a given diffusion tensor in terms of the 
deformation quantities.

In this note we present a novel formulation for the description of soft active 
deformable media within the context of coupled reaction-diffusion-mechanics 
systems, and employ nonlinear cardiac dynamics as a main motivating example. 
At this point we stress that the concept of stress-assisted diffusion has been 
originally formulated for generalised composite media (see~\cite{aifantis:1980,weitsman:1987,klepach:2014,miehe:2014} 
and the references therein), but many resemblances exist with respect to 
the active deformation of soft tissues.  Most notably, here we have found  that an anisotropic and
inhomogeneous diffusivity is naturally induced by mechanical 
deformations,  thus  affecting the nonlinear dynamics of the spatiotemporal
excitation wave. This important fact implies that the present formulation 
can recover and generalise a large number of electromechanical models 
based on basic FitzHugh-Nagumo-type descriptions~\cite{panfilov:2005,aliev:1996}. 
The most relevant additional parameters are here the weights accompanying the stress 
when incorporated into the diffusion tensors, and therefore we study the plausibility of specific 
choices in the model parameter space. 
Our assessment is conducted for stretched tissues,
focusing on appropriate physical indicators as conduction velocity and
propagation patterns, and also carefully identifying conditions leading to the 
stability of the coupled system.

%{At this point we emphasize that a proper modification    
%of the ideas developed specifically for stress-assisted diffusion, and 
%based on physical grounds is required before constructing a generalisation 
%of the well-established cable theory into cardiac dynamics. More precisely, 
%under the assumption (supported by experimental evidence) 
%that diffusion depends on deformation, first principles state 
%that generic second order tensors (and in particular the 
%diffusion) can be 
%Taylor-expanded, or Group-theory-decomposed in terms of the 
%deformation quantities (as often assumed in piezoelectrics or many 
%other active physical media ruled by electrodynamics).} 
%

\begin{figure}[ht]
\begin{center}
\includegraphics[height=0.24\textwidth]{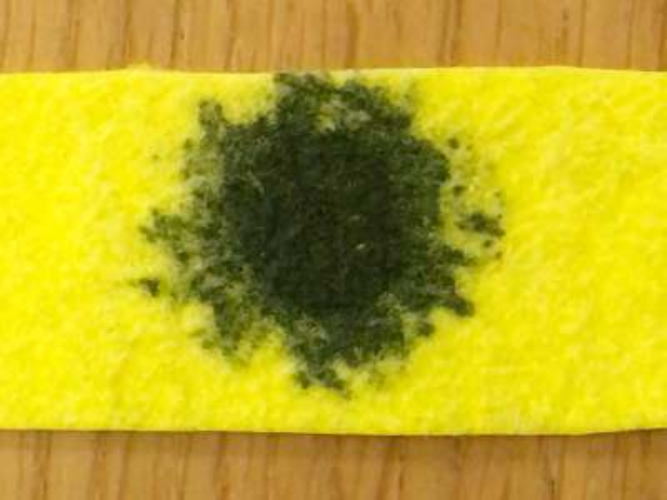}
\includegraphics[height=0.24\textwidth]{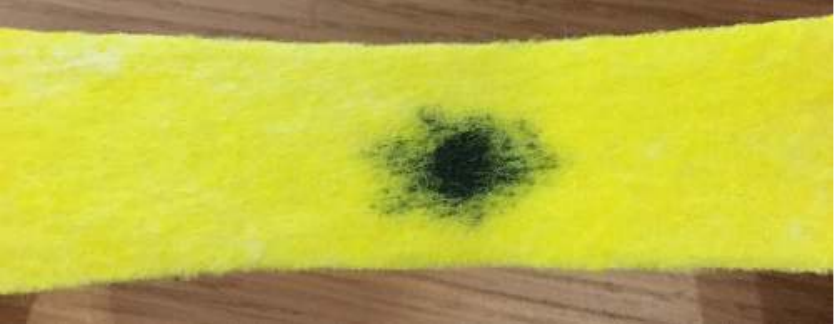}
\end{center}
\caption{Experimental proof of concept. Cleaning sponge non stretched (left) and stretched uniaxially (right). 
A black dye diffuses in the fixed tensile condition, and approximate disk vs. 
ellipsoid shapes are obtained, respectively. Length scale is of the order of cm.\label{fig:exp}}
\end{figure}

Before stating the proposed set of governing equations describing 
the stress-assisted diffusion model, we provide a proof-of-concept, 
elementary experiment. It consists 
of a dye diffusing on a sponge, where stress-assisted diffusion is readily 
encountered by simply applying stretch to the body. 
Figure~\ref{fig:exp} presents two mechanical states of the system, where 
the effects of the underlying 
structural features of the medium (with respect to the expected propagation patterns 
without stretch) are illustrated.   
We do not claim that such a process faithfully represents the physical mechanisms 
ruling biological media, and especially not active materials. Nevertheless, some 
interesting features are indeed shared with the general framework we intend to analyse.

 % ****************************************************************************************

% ****************************************************************************************
\section{A stress-assisted electromechanical model}
We centre our investigation on an active stress RD
model describing non-oscillatory cardiac tissue that supports
stable propagation of excitation waves~\cite{nash:2004,panfilov:2005}. We frame
our modelling into finite elasticity, where one identifies the relationship between material
(reference) and spatial (deformed) coordinates, indicated by $X_I$
and $x_j$, respectively, via the smooth map $x_i(X_I)$ that
determines then the deformed position of a point $x_i$ originally
located at $X_I$.  We indicate with $J$ the Jacobian of the map. 
In the deformed configuration the proposed equations read:
\begin{alignat}{2}
    \dfrac{\partial V}{\partial t} &=
    \dfrac{\partial }{\partial x_i} d_{ij}(\sigma_{ij}) \dfrac{\partial V}{\partial x_j} + I_{\rm ion},
    &\quad
    \dfrac{\partial r}{\partial t} &= f(V,r) \,,
    \label{eq:RD}
    \\
    \dfrac{\partial T_a}{\partial t}
    &= \epsilon(V) (k_{T_a} V - T_a) ,
    &\quad
    \dfrac{\partial \sigma_{ij}}{\partial x_i} &= 0 \,,
    \label{eq:Mecc}
\end{alignat}
with constitutive prescriptions for the RD system~\cite{panfilov:2005}
\begin{subequations}
\begin{align}
    I_{\rm ion}
    &=
    - k V (V-a)(V-1) -rV
    \\
    f(V,r) &=
    \left( \varepsilon + \dfrac{\mu_1 r}{\mu_2 + V}\right) \left( -r-kV(V-b-1)\right)
\end{align}
\label{eq:RDf}
\end{subequations}
and for incompressible isotropic materials $(J=1)$
\begin{eqnarray}
    \sigma_{ij} &=& 2 c_1 b_{ij} - 2 c_2 b^{-1}_{ij} - p \delta_{ij}+ 
    T_a \delta_{ij} \,,
    \label{eq:Stress}
    \\
    d_{ij}(\sigma_{ij}) &=& D_0 \left( \delta_{ij} + D_1 \sigma_{ij} + D_2 \sigma_{ik}\sigma_{kj} \right)\,.
    \label{eq:Diffusivity}
\end{eqnarray}
Equation \eqref{eq:RD} provides a non-dimensional, two-variables RD
model where $V$ here represents the transmembrane potential and $r$ is
the recovery variable, whose dynamics is prescribed by
Eqs.~\eqref{eq:RDf}. Equation \eqref{eq:Mecc} provides the contractile
(active stress) dynamics of the tissue and the balance of linear
momentum (law of motion). Here $\epsilon(V)$ is a Heaviside function
introducing a discontinuous coupling (a switch) with the dynamics of
the voltage~\cite{nash:2004}: $\epsilon(V) = \epsilon_0$ if $V<0.05$, and $\epsilon(V)
= 10\epsilon_0$ if $V\ge0.05$. Equation \eqref{eq:Stress} provides the
total equilibrium stress in the current deformed configuration
($\sigma_{ij}$ is the Cauchy stress~\cite{spencer:1980}). Such a
formulation highlights the two main contributions due to the
multiscale behaviour of the tissue, i.e. (i) the passive stress
component  originating from the classical strain energy function adopted
for incompressible Mooney-Rivlin materials~\cite{ogden} and (ii) the active stress
component following the formulation proposed in \cite{nash:2004}, 
contributing as an additional hydrostatic term to the total stress
tensor only.  The second order tensors $\delta_{ij}$ and $b_{ij}$ are
the identity and left Cauchy-Green strain~\cite{spencer:1980},
respectively. Model parameters are provided in Tab.~\ref{table}.

Equation \eqref{eq:Diffusivity} introduces the stress-assisted
diffusion equation \cite{aifantis:1980} generalising the effects of
tissue deformation on action potential spread for isotropic and
incompressible nonlinear elastic solids. 
On these assumptions, and according to the
representation formula for isotropic second order tensors
\cite{spencer:1980}, the diffusive flux accounts for hydrostatic
stress effects on the diffusing species and is characterised by three
contributions: $D_0 \delta_{ij}$ recovering the classical linear
diffusion equation~\cite{nash:2004}, 
 while $D_1 \sigma_{ij}$ and $D_2 \sigma_{ik}\sigma_{kj}$ allow for
stress-induced anisotropies into the diffusion tensor (this part is
associated to the theory by Landau et al.~\cite{landau:1984} for
infinitesimal deformations). 
We disregard here stretch-activated currents, as introduced later 
in~\cite{panfilov:2005}, and follow instead the treatment in 
\cite{nash:2004} to highlight the new main ingredients of the work. 

%**************
\begin{table}[h]
 \begin{center}
\begin{tabular}{c}
\hline\noalign{\smallskip}\hline\noalign{\smallskip}
$ \Omega = (0,250)^2$,  $a=0.1$, $b=0.15$, $k=8$, $\epsilon =0.001$, $\epsilon_0=0.1$, \\
$\mu_1=0.12$, $\mu_2=0.3$, $k_{Ta} = 9.58$, $c_1=6$, $c_2=2$, $D_0=1$, \\ 
 $V_{\text{init}} =0$, $r_{\text{init}} =0.4$, $(T_a)_{\text{init}} = 0.2$, $D_1,D_2$ from Figure~\ref{fig:1a}\\
\hline\noalign{\smallskip}\hline
 \end{tabular}
 \end{center}

\vspace{-3mm}
  \caption{\label{table} Model parameters specifying the phenomenological RD cardiac dynamics and the active-stress mechanics \eqref{eq:RD}-\eqref{eq:Diffusivity}.}
 \end{table}
%
% ****************************************************************************************
\section{Numerical results}
\paragraph{Numerical method.} 
The discretisation of \eqref{eq:RD}-\eqref{eq:Diffusivity} follows the
mixed-primal finite element formulation proposed in \cite{ruiz:2015}, where 
strains and stresses are computed directly without postprocessing them 
from low-order discrete displacements.  
This approach is preferable in the present context as we aim at analysing properties 
of the stress and diffusion tensors directly. Consequently, stresses are approximated
with row-wise lowest order Raviart-Thomas elements~\cite{Raviart1977}, displacements with stabilised Brezzi-Douglas-Marini elements of degree one~\cite{Brezzi1985}, whereas pressure (if actually needed) can be recovered 
from the discrete stress.  The remaining fields (transmembrane potential, recovery variable, and active
tension) are approximated with continuous, piecewise linear elements.
We employ unstructured triangular meshes of 
different resolutions, and 
we set a fixed time step $\Delta t= 10^{-3}$.  The solver
consists of an outer time advancing loop (of first order backward
Euler type), an embedded fixed point iteration to decouple the
electrophysiology RD system from the finite elasticity discrete 
equations, and inner Newton steps for the solution of the nonlinear 
mechanics. In turn, the AP model \eqref{eq:RD} is solved semi-implicitly (taking only the reaction terms explicitly), 
determining a formal CFL condition associated to the finest mesh and 
supporting the choice of the timestep mentioned above.
All linear solves are carried out with a PETSc built-in LU solver using ILU factorisation as 
preconditioner.

\begin{figure}[t!]
\begin{center}
\includegraphics[width=0.325\textwidth]{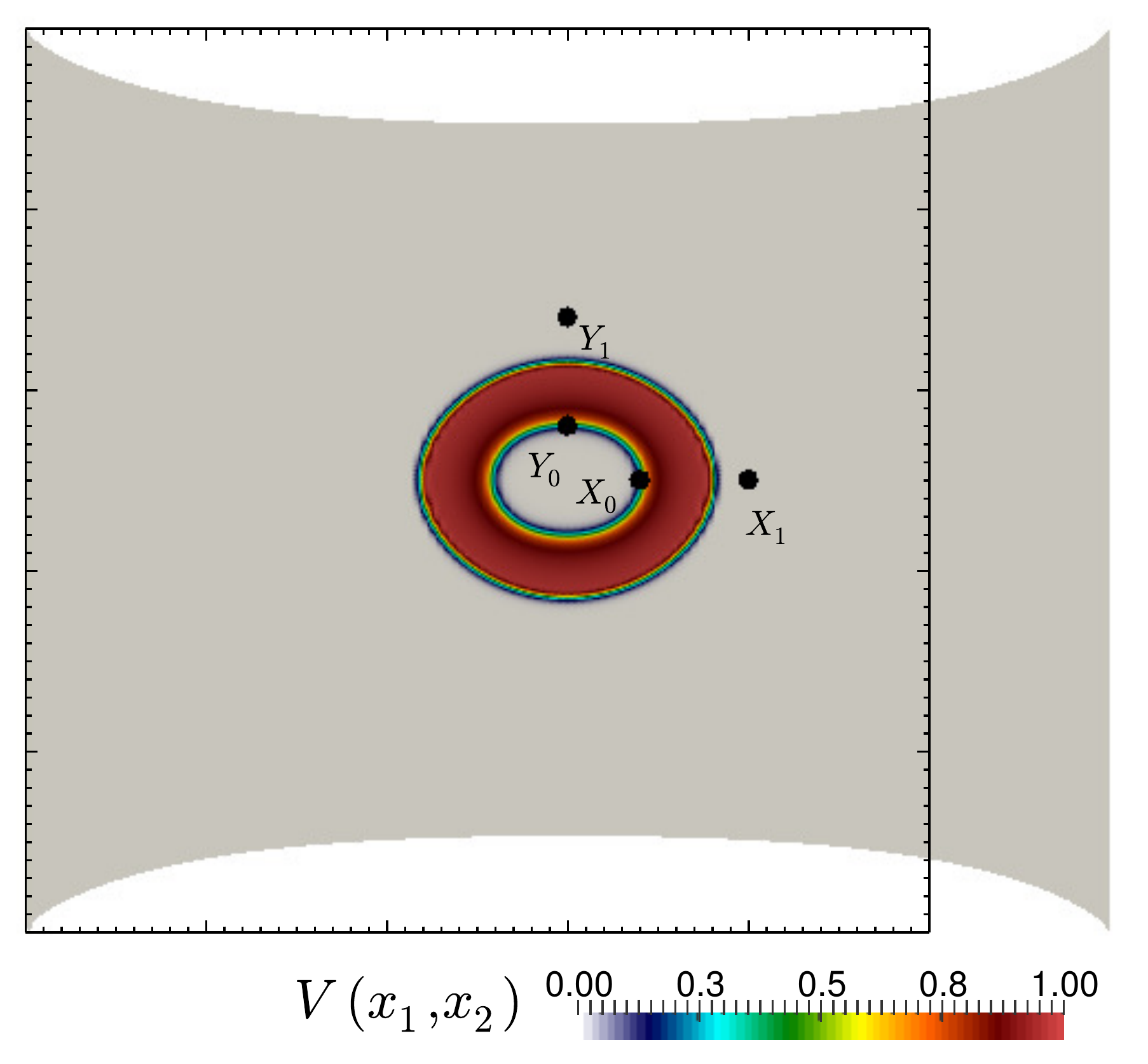}
\includegraphics[width=0.325\textwidth]{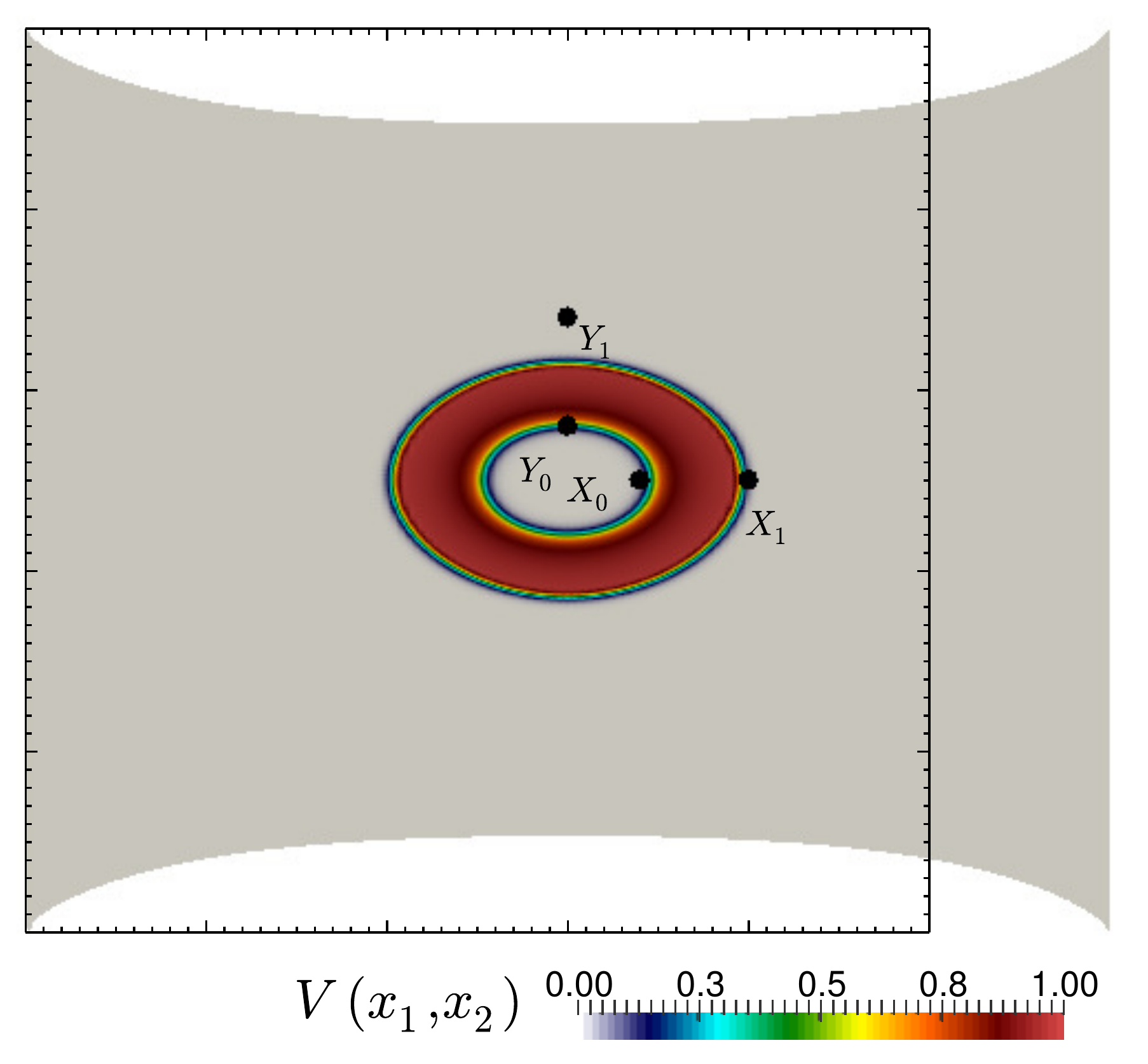}
\includegraphics[width=0.325\textwidth]{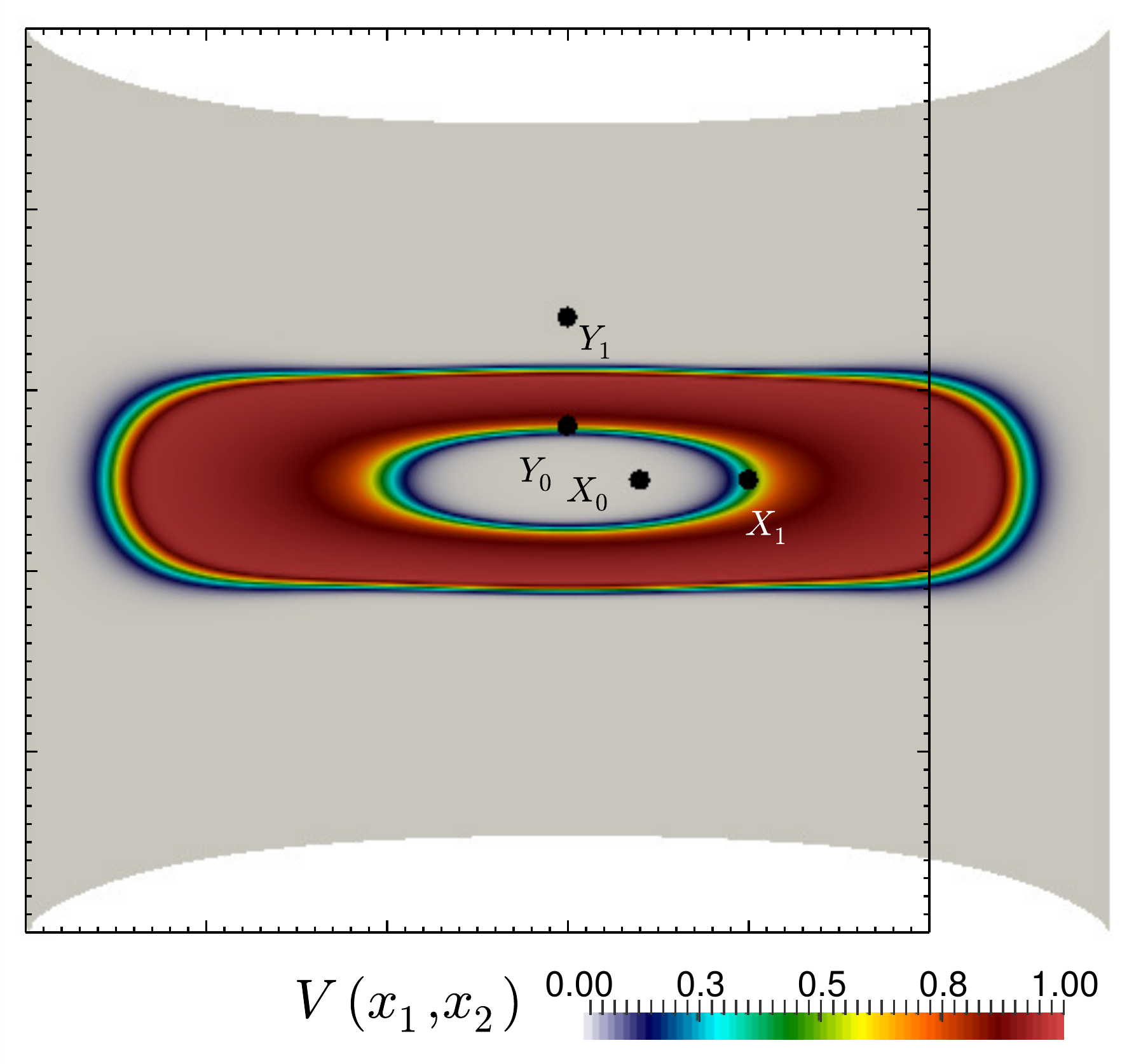}\\
\subfigure[]{\includegraphics[width=0.325\textwidth]{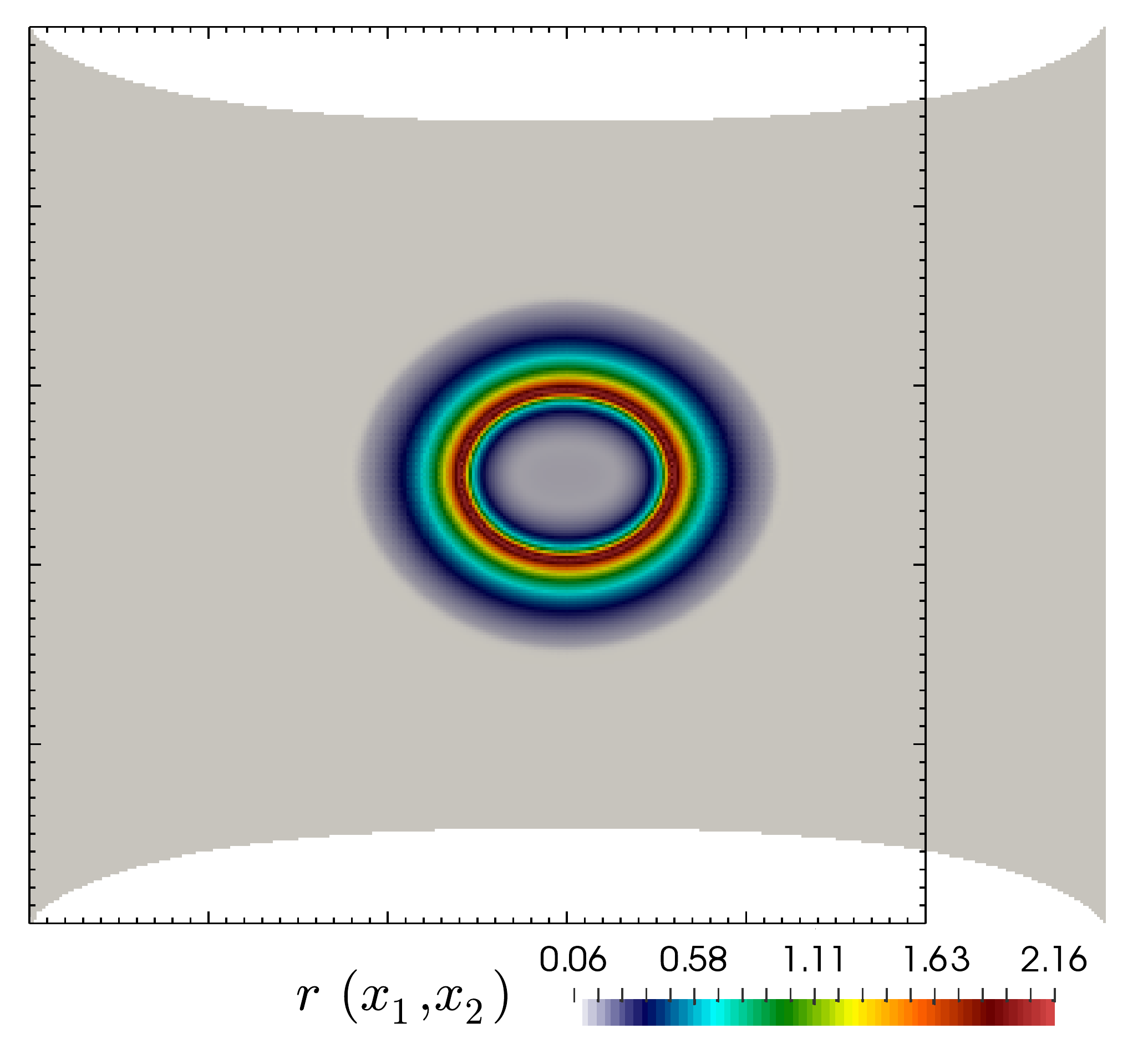}\label{fig:1Ba}}
\subfigure[]{\includegraphics[width=0.325\textwidth]{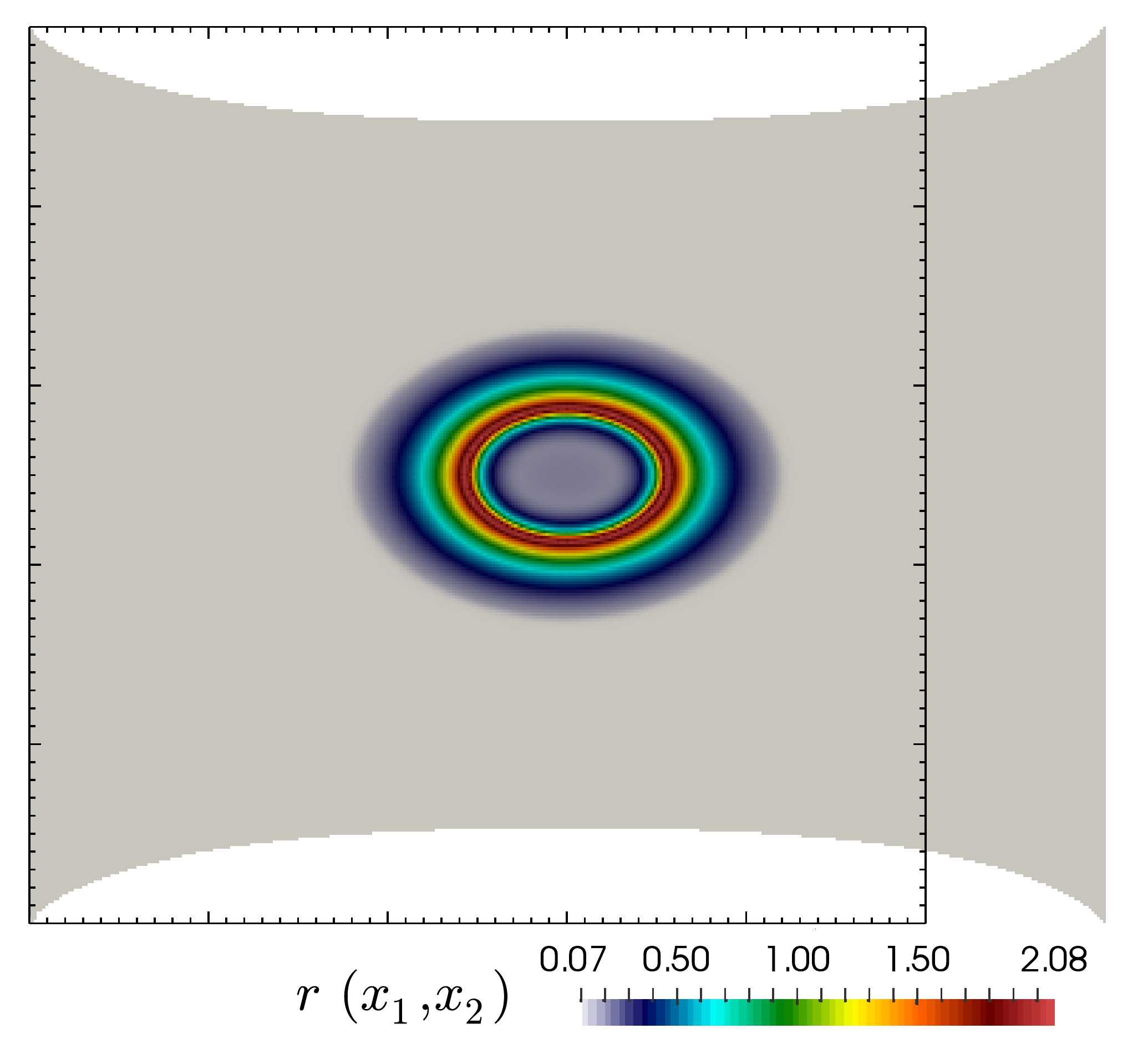}\label{fig:1Bb}}
\subfigure[]{\includegraphics[width=0.325\textwidth]{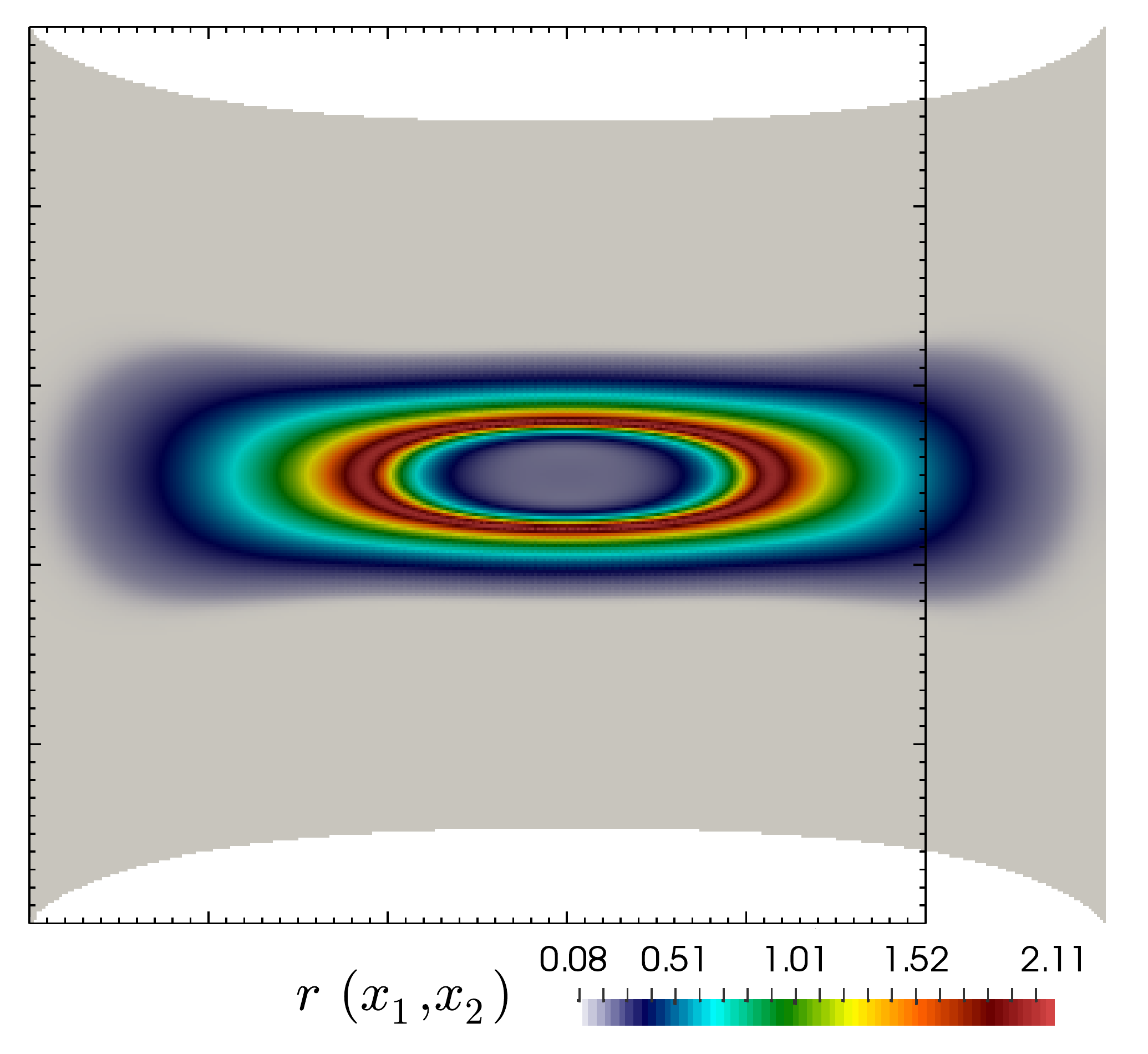}\label{fig:1Bc}}
\end{center}
\caption{Stress-induced anisotropy. Target pattern
  simulation with $D_1=D_2=0$ according to \cite{nash:2004} (a),
  $D_1=-0.01,D_2=0.01$ (b) and $D_1=-0.25,D_2=0.25$
  (c). Locations ($X_0,Y_0,X_1,Y_1$) refer to the horizontal and vertical directions used to compute the corresponding conduction velocities $CV_x, CV_y$. Top panels show the electric potential and the bottom row presents the distribution of the recovery variable.\label{fig:1BB}}
\end{figure}

% ****************************************************************************************
\paragraph{Parameter space.} 
In order to determine admissibility regions for the modified RD
system, we conduct an inspection of the ellipticity regimes depending
on the model parameters (see e.g.~\cite{showalter}).  Let us first consider a uniaxial tensile
configuration of a two-dimensional domain where we induce a target
pattern (see Figure~\ref{fig:1BB}, left) and identify horizontal ($X_0,X_1$) and vertical ($Y_0,Y_1$) positions.  Fixing the stretching of
the domain at the maximum physiological level for cardiac tissue,
i.e. 20\% of the resting length, one can obtain the maximum tensile
components and a diagonal form of the Cauchy stress tensor can be
readily derived in the central uniform tensile region~\cite{ogden},
i.e.:
$$
	[\sigma_{ij}] = 
	\left[
	\begin{matrix}
		\sigma_1 & 0 \\
		0 & \sigma_2
	\end{matrix}
	\right],
	\qquad
	[\sigma_{ik}\sigma_{kj}] = 
	\left[
	\begin{matrix}
		\sigma_1^2 & 0 \\
		0 & \sigma_2^2
	\end{matrix}
	\right],
$$
where $\sigma_1,\sigma_2$ correspond to the eigenvalues of the stress
tensor that overlap with the two diagonal components of the stress in
the Cartesian coordinate system, i.e., $\sigma_{11}, \sigma_{22}$. Accordingly,
the generalised diffusion tensor \eqref{eq:Diffusivity} becomes
$$ [d_{ij}] = 
	\left[\begin{matrix}
		1 + D_1 \sigma_1 + D_2 \sigma_1^2 & 0 \\
		0 & 1 + D_1 \sigma_2 + D_2 \sigma_2^2
	\end{matrix}
	\right] \,, 
$$
where a second order polynomial expression is obtained and parametrised
over the two coefficients $D_1,D_2$. We can then readily identify the
regimes where the modified conductivity is a positive definite tensor.
This translates in requiring that the graph of $y=1+D_1\sigma+D_2\sigma^2$
is always positive, i.e., 
$${\frac{-D_1 \pm \sqrt{D_1^2 -4D_2}}{2D_2} \ge 0\,, \quad\text{and}\quad D_1 \ge \pm 2\sqrt{|D_2|} \,.}$$
In particular, this condition ensures that both
tensile and compressive states are physically allowed according to the
material response without violating ellipticity.
%

%****************
\begin{figure}[t!]
\begin{center}
\subfigure[]{\includegraphics[width=0.325\textwidth]{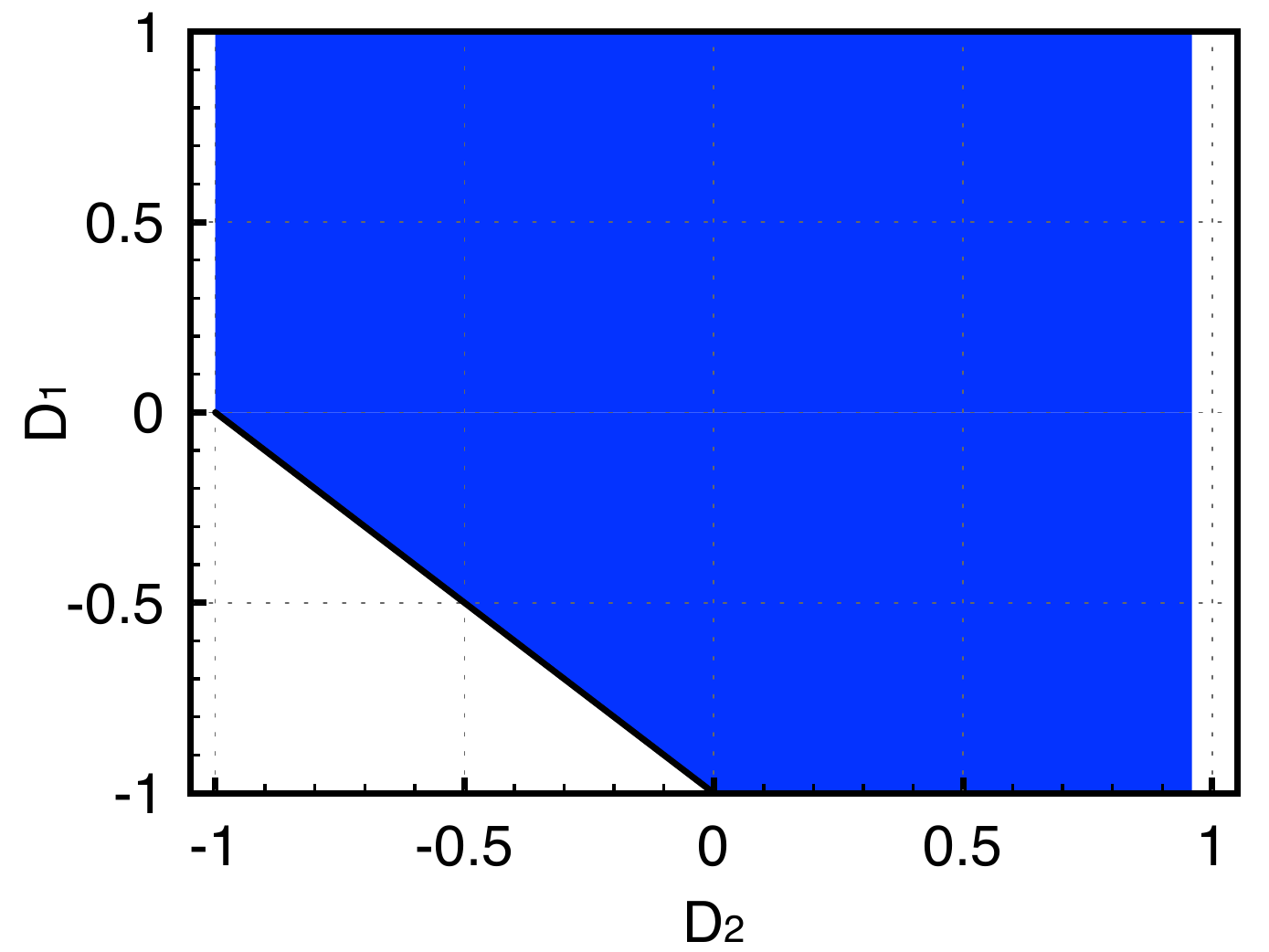}\label{fig:1a}}
\subfigure[]{\includegraphics[width=0.325\textwidth]{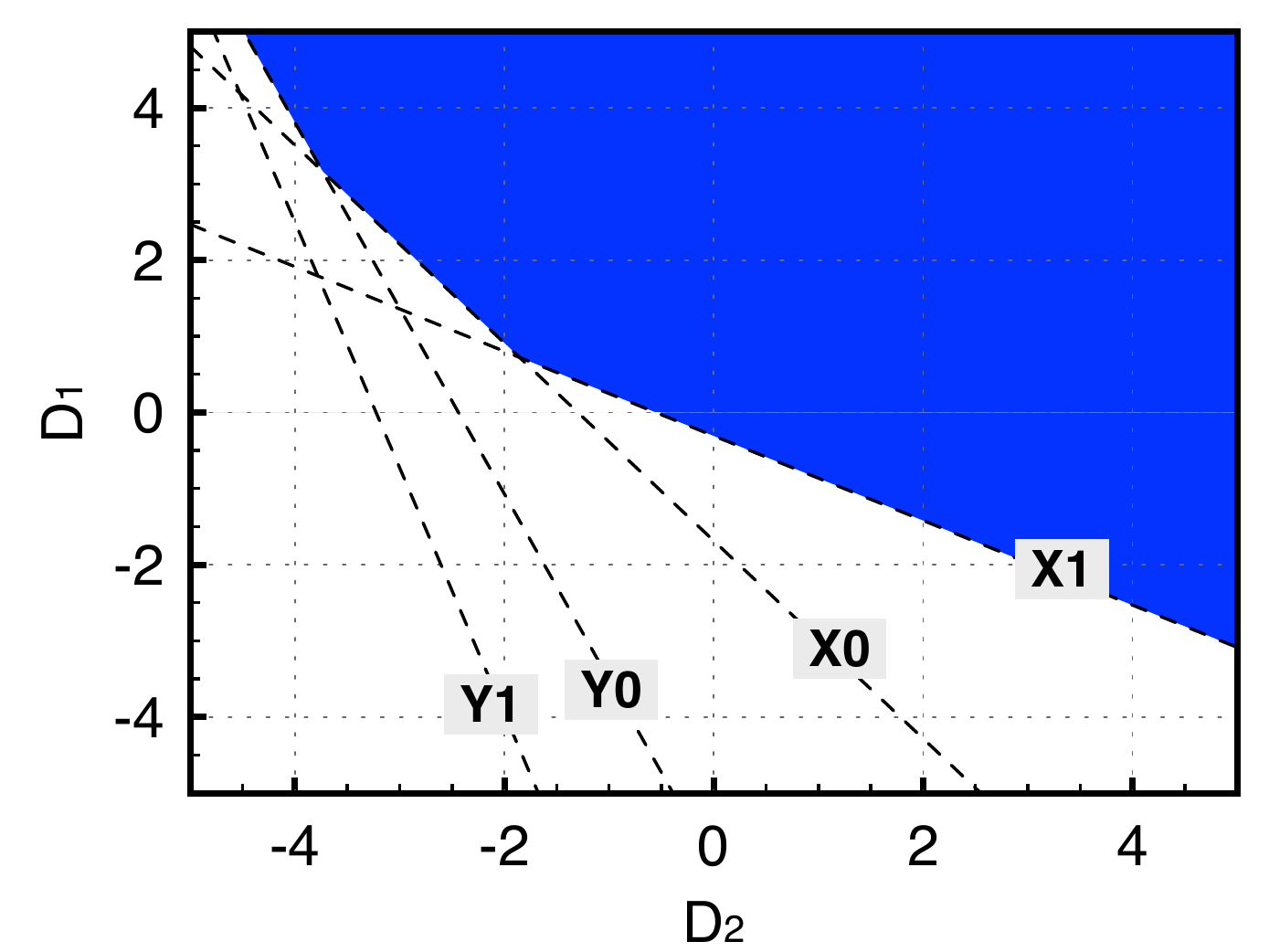}\label{fig:1b}}
\subfigure[]{\includegraphics[width=0.325\textwidth]{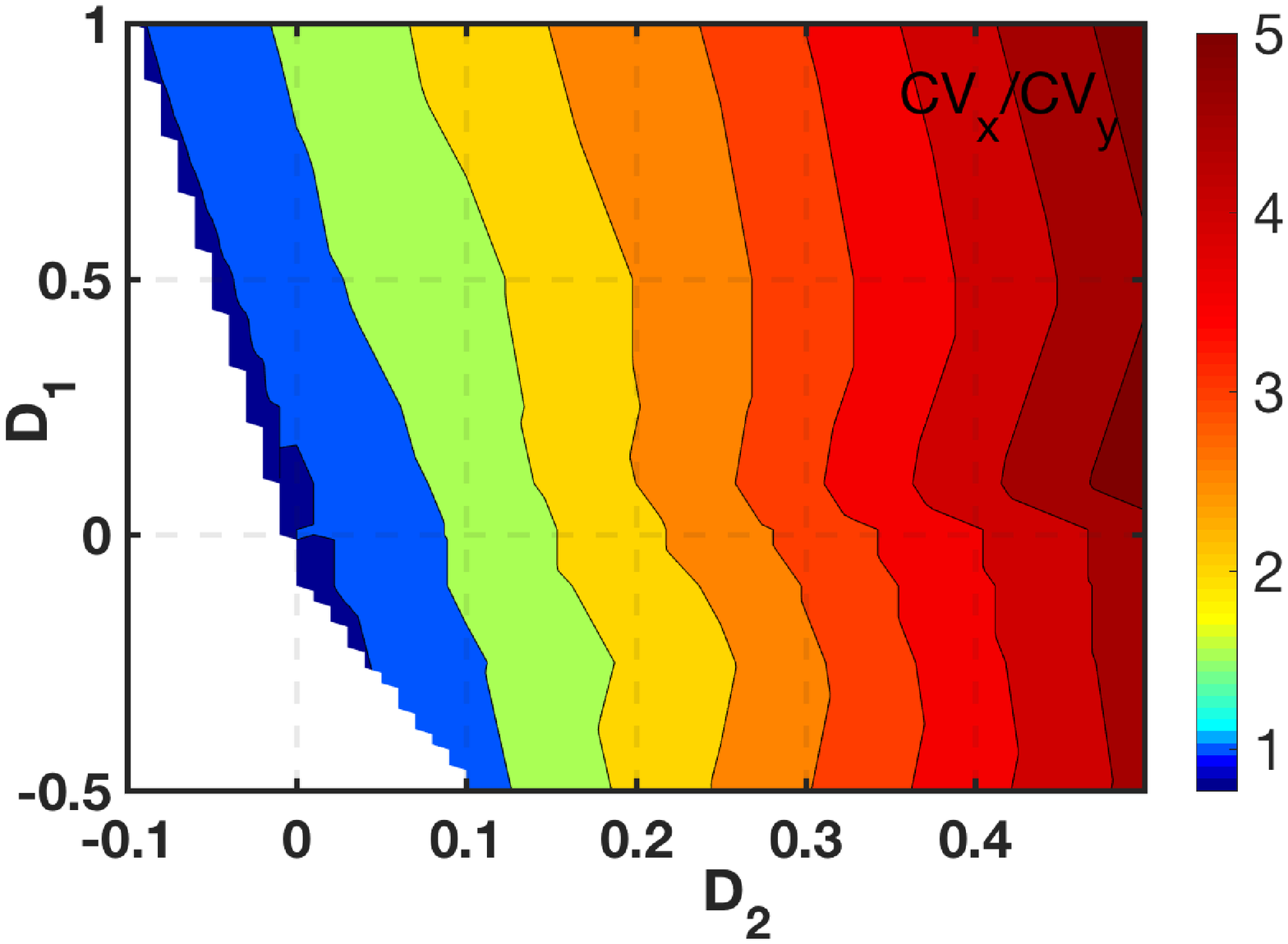}\label{fig:1c}}
\end{center}

\vspace{-3mm}
\caption{Parameter space. Conditions for ellipticity of the
  generalised diffusion tensor \eqref{eq:Diffusivity}:
  (a) theoretical normalised case for unitary stress
  ($\sigma_1=\sigma_2=1$); 
  (b) computed stress values
  at locations $X_0,X_1,Y_0,Y_1$ during uniaxial loading, 
  (c) numerical conduction velocity ratio between horizontal
  $CV_{x}$, ($X_0,X_1$ locations), and vertical $CV_{y}$,
  ($Y_0,Y_1$ locations), computed during target pattern
  stimulation in Figure~\ref{fig:1BB}.\label{fig:1AA}}
\end{figure}

We proceed to characterise the theoretical parameter space for an
idealised tensile state $(\sigma_1=\sigma_2=1)$. Figure~\ref{fig:1a} 
displays the map  $f(D_1,D_2)=1+D_1/D_2 + 1/D_2$, highlighting in blue
colour the admissible regions where ellipticity holds. On the other
hand, we perform numerical simulations (with stresses $\sigma_1,\sigma_2$) and extract the computed values
of the stresses at selected locations ($X_0,Y_0,X_1,Y_1$) whose
ellipticity conditions are shown in Figure~\ref{fig:1b}.
Finally, the physical parameter space is derived directly from
numerical simulations for the same uniaxial condition in
Figure~\ref{fig:1c}. We compute the conduction velocities of
the generated wave along the selected locations, $CV_x,CV_y$, and identify regions
with loss of ellipticity when the travelling waves present a
propagation fault (non coloured region). 
In particular, the produced values of $CV_x/CV_y$ are for the most larger than 1 
since the mechanical stretching is applied in the $x-$direction 
and the computations are non-diverging for most positive parameter values.
Close to the region $D_1=D_2=0$, the CV ratio is $\sim1$, thus recovering 
the conduction velocity expected for an isotropic medium (that is, 
the same independently of the direction).
 Particular combinations with $D_1$ or $D_2<0$
are characterised by $CV_x/CV_y<1$ thus implying a reduction of the 
conduction velocity in the direction of stretching (see regions in blue, in Figure~\ref{fig:1c}). 
In contrast, for larger positive parameter values 
the wave moving in the $x-$direction propagates 
much faster than the one in the $y-$direction (up to a fivefold in our analyses). 
In all the selected scenarios, the analysis 
clearly indicates a marked anisotropic effect on the diffusion process, due to 
stress. 

%THIS PARAGRAPH IS NOT CLEAR. I've JUST REMOVED IT
%In this elementary setting, the problem behaves symmetrically if mechanical stretch is applied in the 
%vertical direction,} {\red however, in general, interesting phenomena can be obtained 
%also in such a minimal scenario as described below. [not clear]
%}

\paragraph{Stress-induced anisotropy.}
Figure~\ref{fig:1BB} 
 shows three representative examples of different levels of the
stress-assisted anisotropic coupling. One can observe how the  
propagating behaviour is modified by the level of anisotropy induced on the tissue, 
initially homogeneous and isotropic, thanks to the stress-assisted mechanism.
Based on the selected parameters, 
a faster propagation is obtained in the direction of loading where a horizontal
uniaxial stress state is induced. In the first case (corresponding to 
Figure~\ref{fig:1Ba}), the choice $D_1=D_2=0$ does not induce
any effect on the excitation wave and the only visual difference is due to
the geometric mapping from the reference to the deformed configuration of the
computational domain.  
Actually, this setting (which is the expected behaviour in the classical active stress 
approach for incompressible isotropic materials~\cite{nash:2004}) 
is the most commonly employed in 
models and simulations for the electromechanical coupling. \\ 
The second case (see Figure~\ref{fig:1Bb})  is characterised by
$D_1=-0.01,D_2=0.01$ and shows a mild, though significative, anisotropic effect
due to the applied mechanical stretching in the horizontal direction.
The obtained behaviour is the one expected based on the few cardiac 
optical mapping data describing such a phenomenon 
in a moving frame~\cite{miller:2002} and, out of these three cases, 
it is qualitatively the most appropriate, 
in terms of physiological behaviour of cardiac dynamics.
\\
The last run (see Figure~\ref{fig:1Bc}) uses the values $D_1=-0.25,D_2=0.25$, 
and our results produce a $CV_x$ about a threefold higher than $CV_y$. 
These tests point out a critical example of the effect of mechanical stretching on the diffusive 
properties of the species in terms of transmembrane potential $V$ (top panels) and
recovery variable $r$ (lower panels). 
The observed level of distortion induced on the excitation wave may be expected
in some pathological scenarios with drastically compromised excitation-contraction dynamics~\cite{orourke:1999}. 
We remark that this analysis is of particular importance 
in the study of the effects of the electromechanical coupling into 
spiral wave drifting phenomena and atria arrhythmias.

\begin{figure}[t!]
\begin{center}
\includegraphics[width=0.325\textwidth]{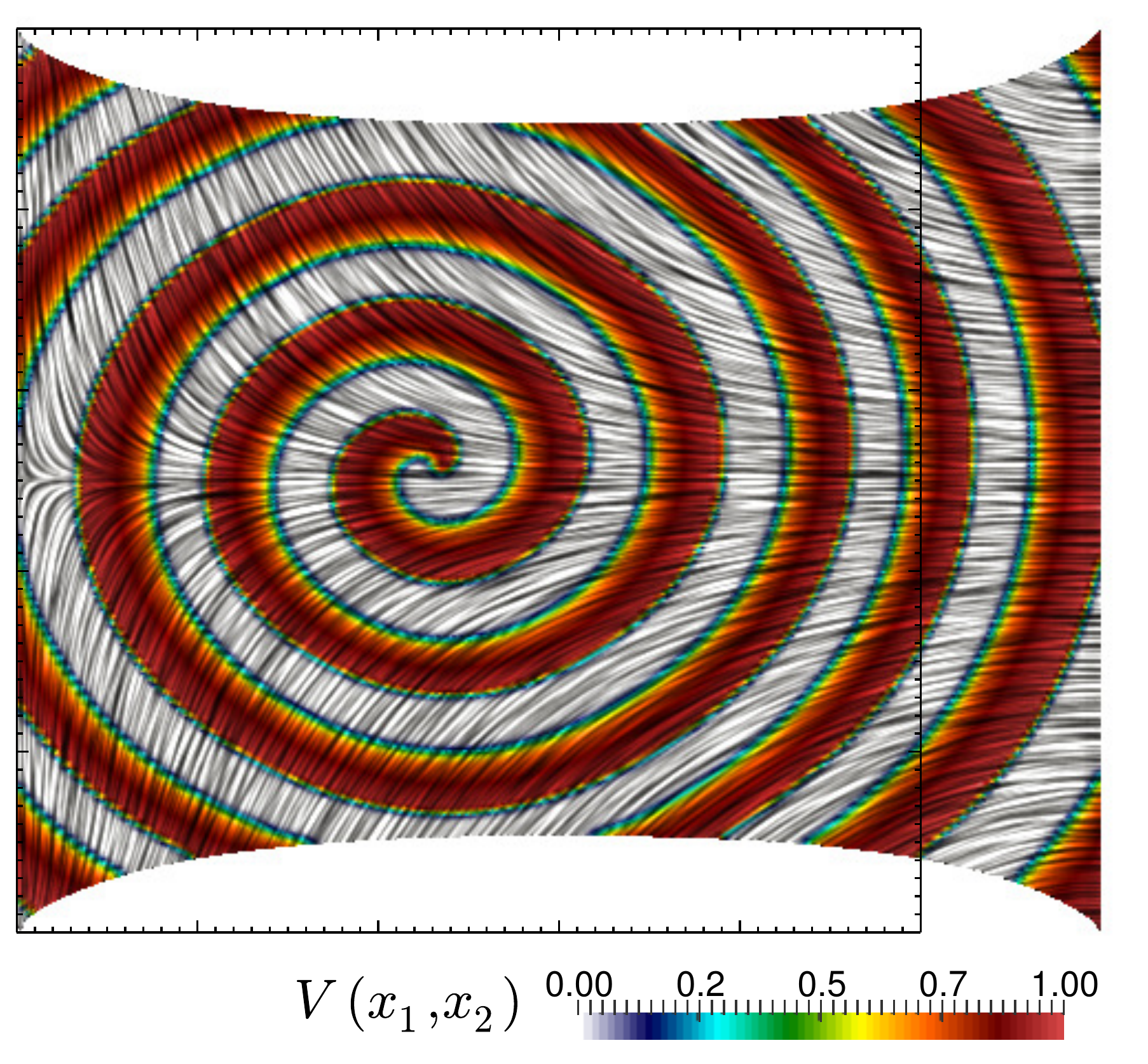}
\includegraphics[width=0.325\textwidth]{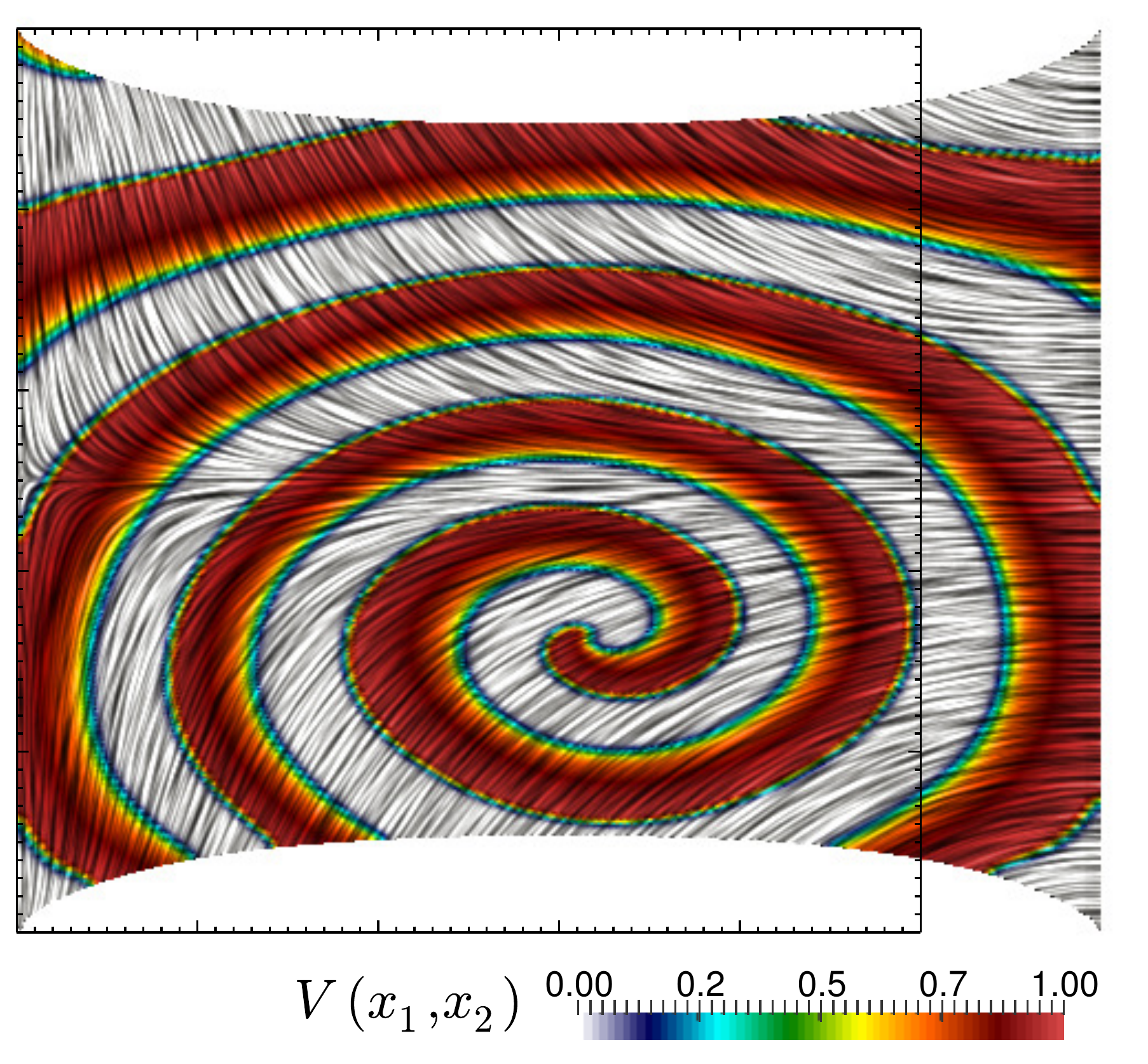}
\includegraphics[width=0.325\textwidth]{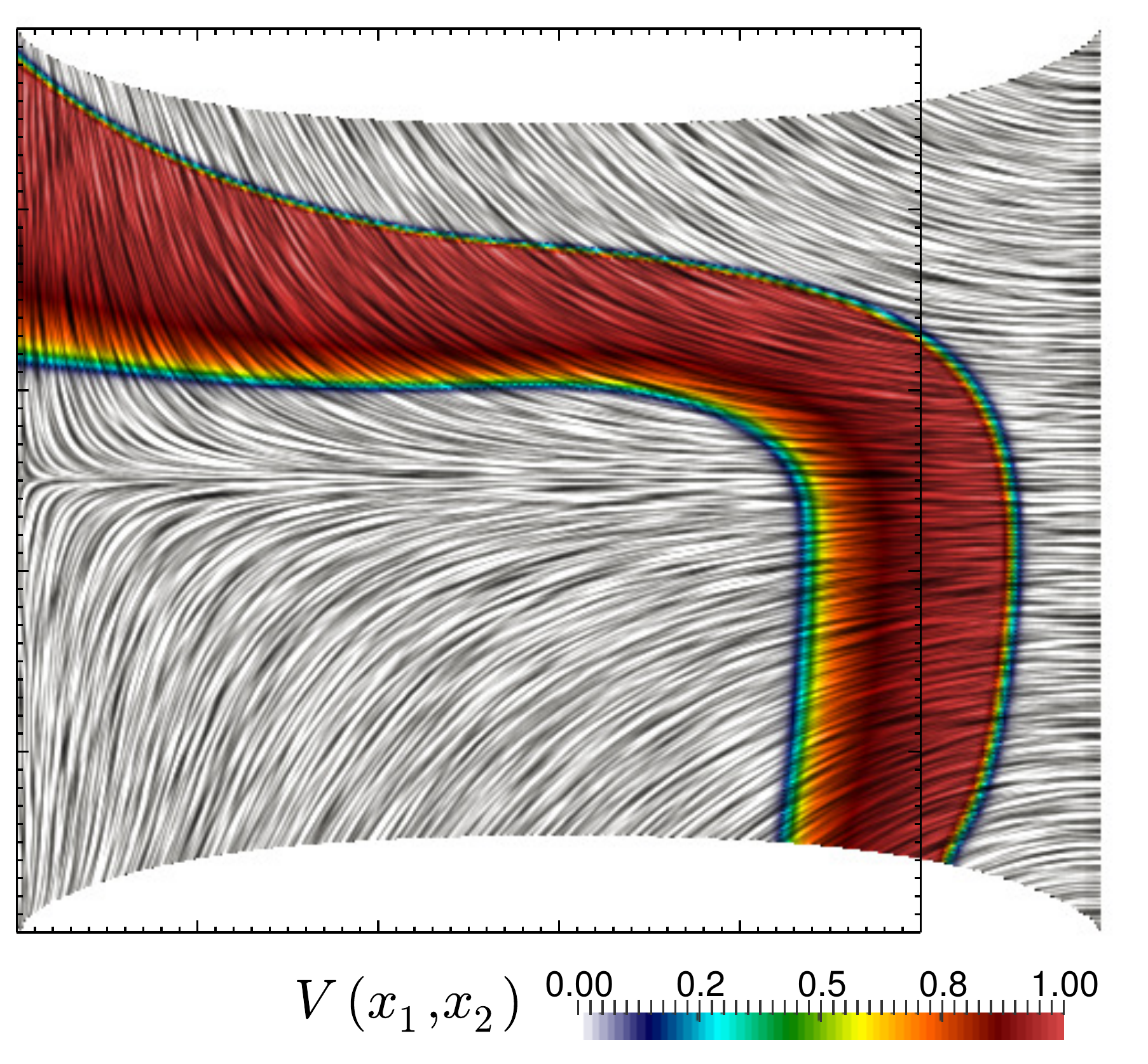}
\\
\subfigure[]{\includegraphics[width=0.325\textwidth]{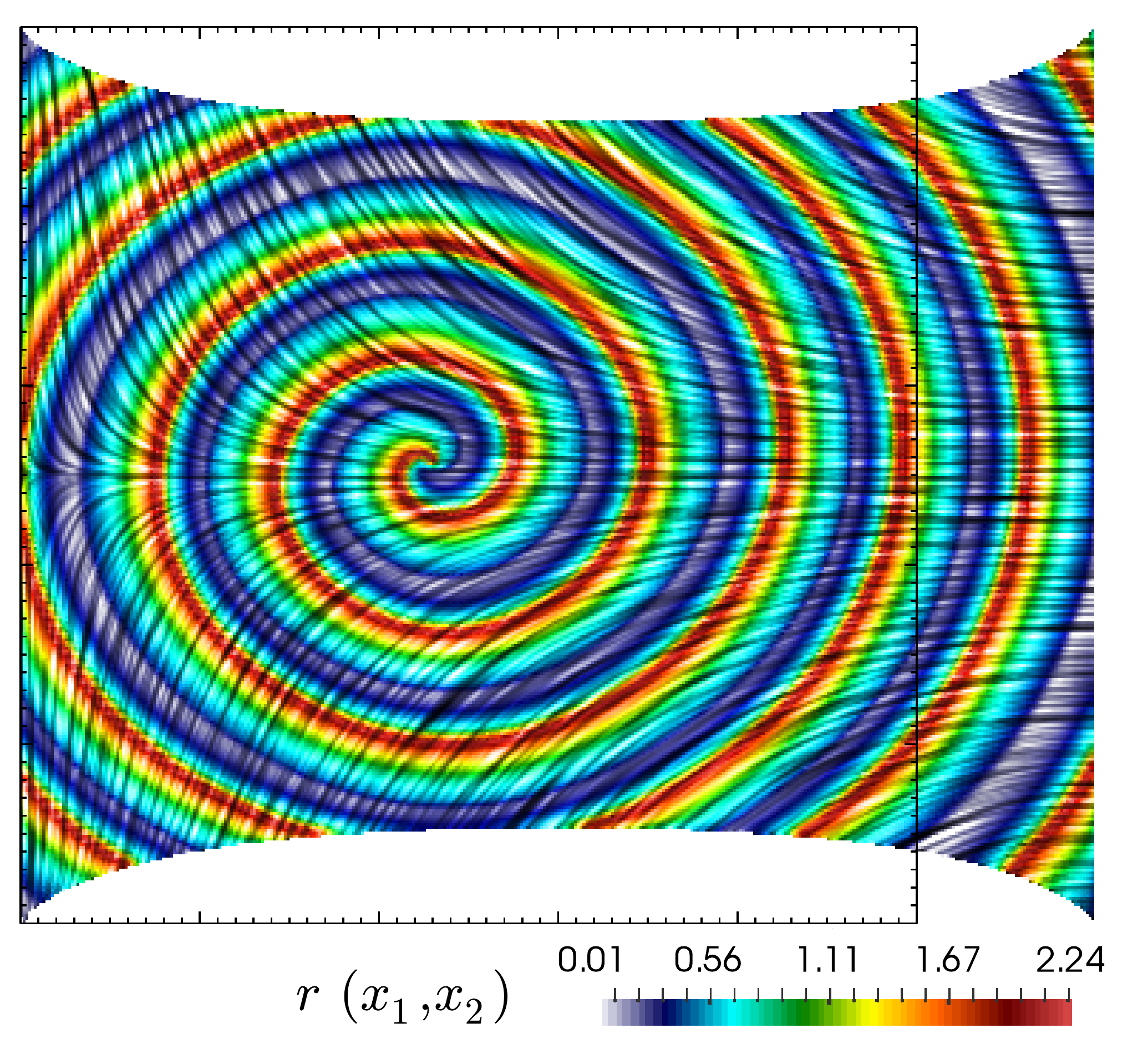}\label{fig:4a}}
\subfigure[]{\includegraphics[width=0.325\textwidth]{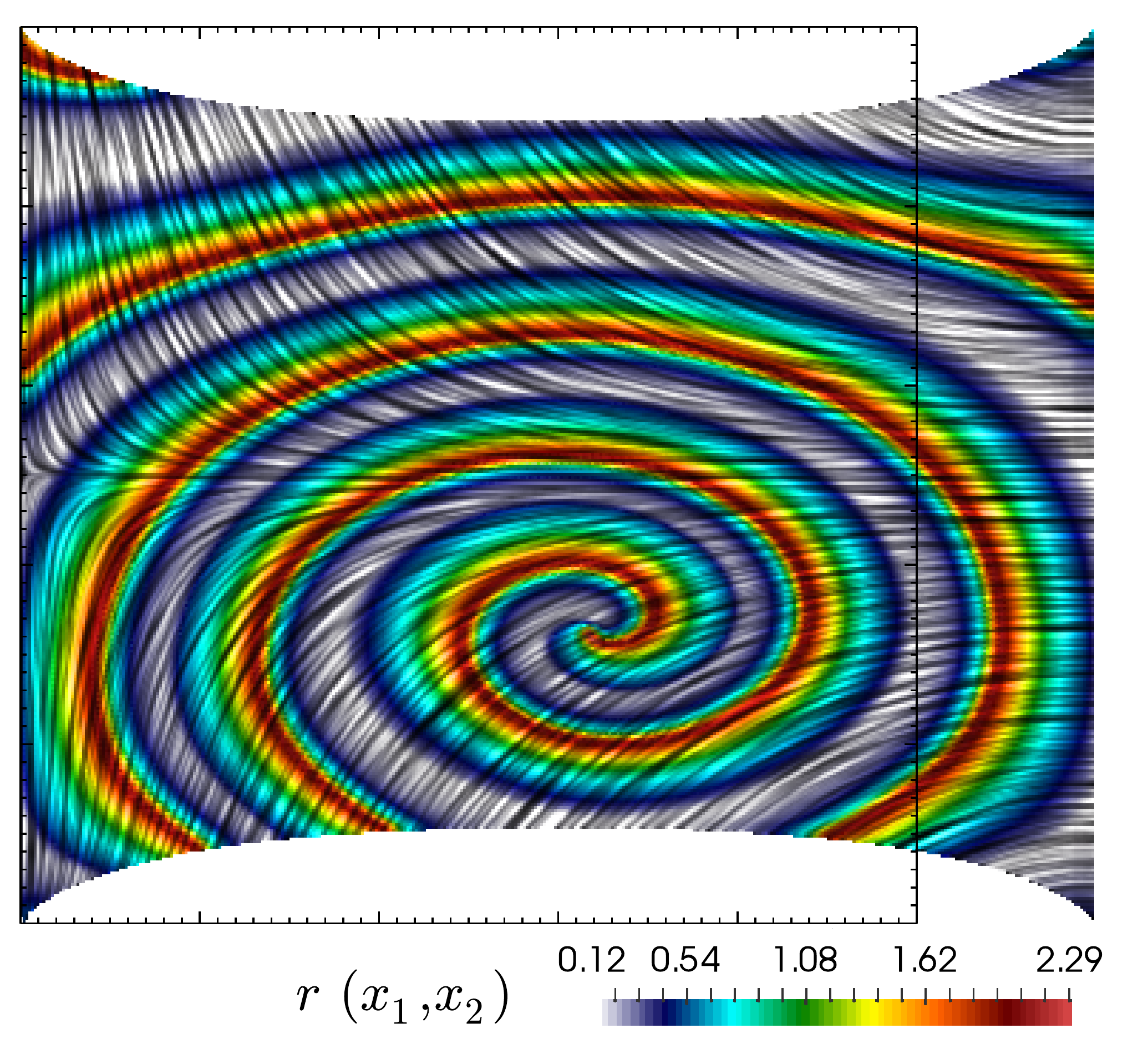}\label{fig:4b}}
\subfigure[]{\includegraphics[width=0.325\textwidth]{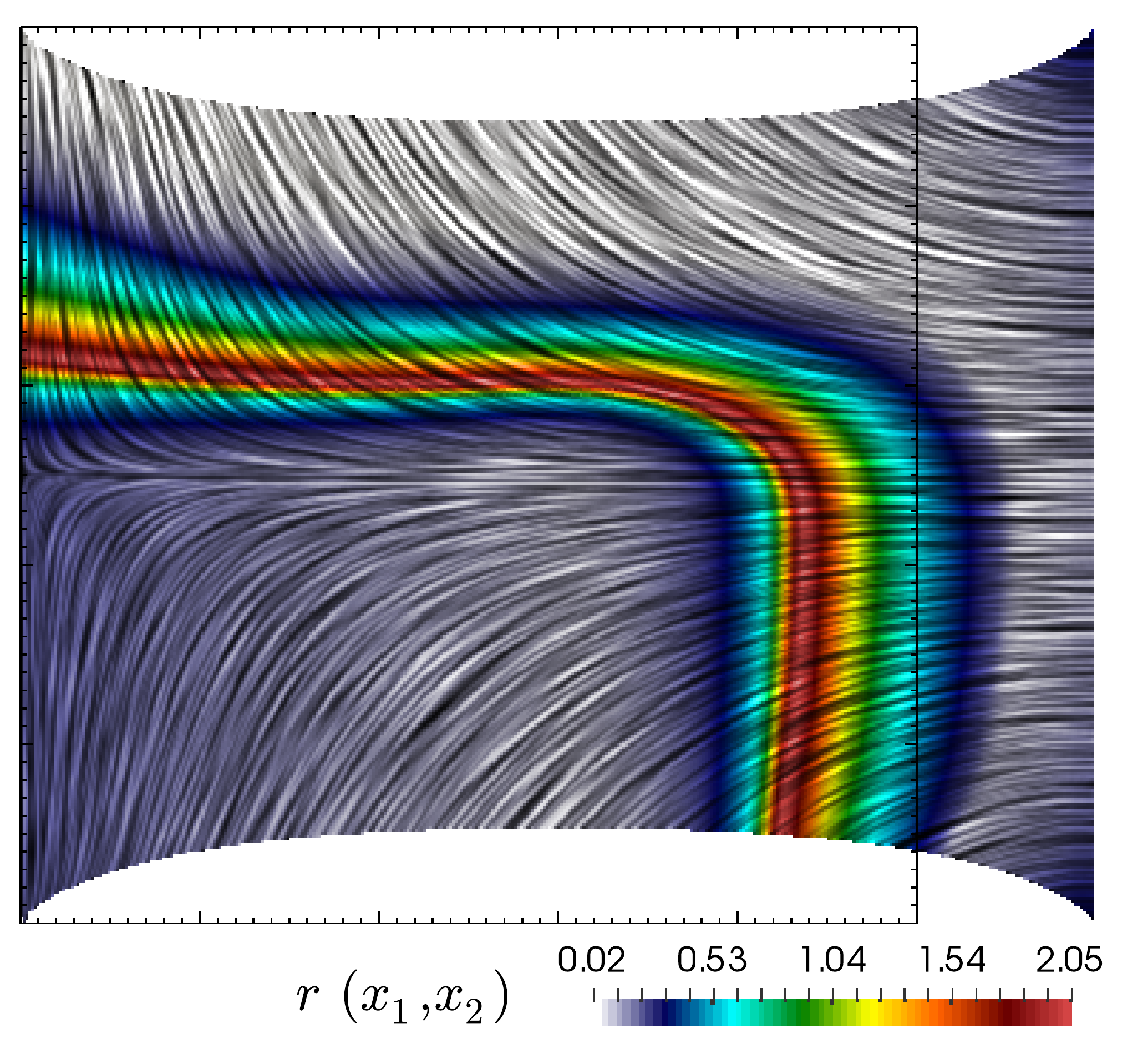}\label{fig:4c}}
\end{center}
\vspace{-3mm}
\caption{Stress-induced arrhythmogenesis. Comparison of spiral dynamics induced by (a) constant diffusion according to \cite{nash:2004}
  and to Piola transformation (i.e. $D_0C_{IJ}^{-1}$), and those
  generated by stress-assisted diffusion with $D_1=-0.015,D_2=0.015$
  (b) and $D_1=0.15,D_2=0.015$ (c).
  Top: Action potential shape distribution $(V)$.
  Bottom: recovery variable distribution $(r)$.\label{fig:4}}
\end{figure}

%******************************************** 
\paragraph{Stress-induced arrhythmogenesis.}
Additional effects on arrhythmogenesis are exemplified with a simulation of sustained spiral
dynamics. Figure \ref{fig:4} compares the long run behaviour of an
equally induced spiral within a uniaxially stretched tissue for three
increasing levels of the stress-assisted diffusion parameters $D_1$
and $D_2$. 
For the chosen parameters' set from the original model
a stationary circular meander pattern is expected (Figure~\ref{fig:4a}), 
although spiral drifting is observed as the stress-assisted diffusion is activated (Figure~\ref{fig:4b}). 
In addition, totally distorted excitation waves are observed when a strong anisotropic coupling is enforced (Figure~\ref{fig:4c}) via higher values of the parameters $D_1$ and $D_2$. The corresponding recovery variable is also provided and confirms the highlighted differences. 

\begin{figure}[t!]
 \begin{center}
\subfigure[]{\includegraphics[width=0.325\textwidth]{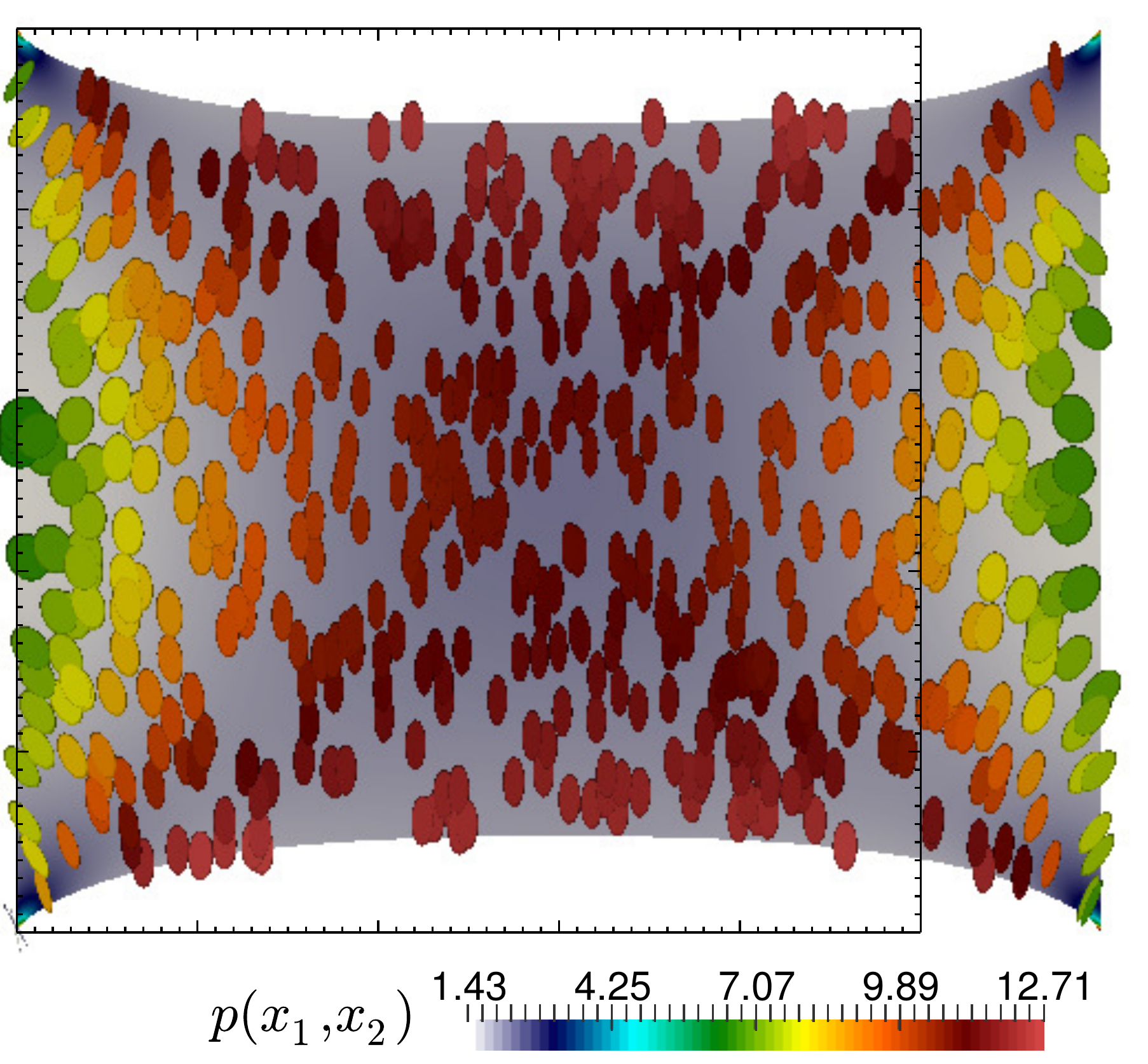}\label{fig:5a}}
\subfigure[]{\includegraphics[width=0.325\textwidth]{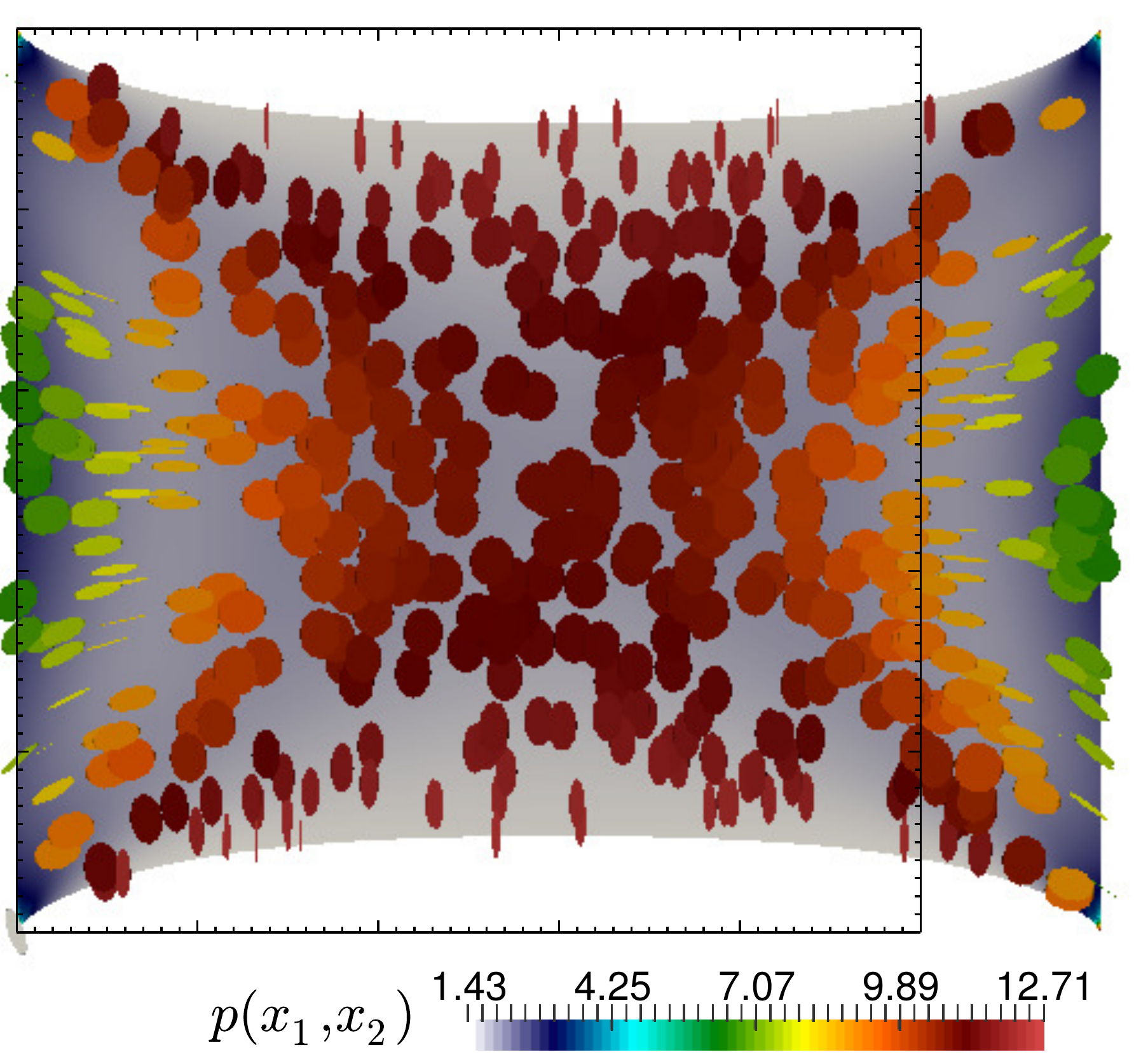}\label{fig:5b}}
\subfigure[]{\includegraphics[width=0.325\textwidth]{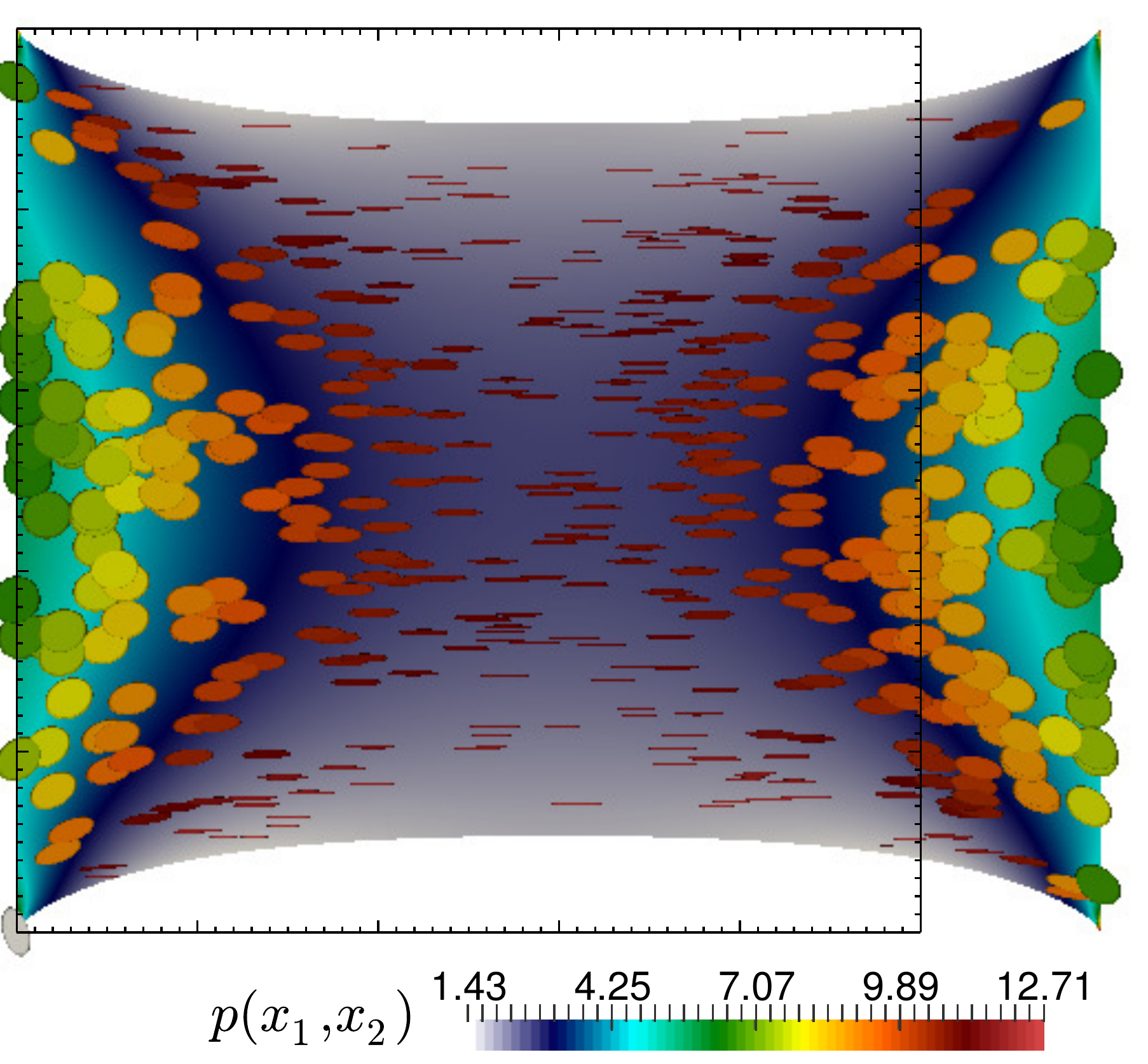}\label{fig:5c}}
\end{center}
\vspace{-3mm}
\caption{Stress-induced arrhythmogenesis. Tensorial glyphs representation with pressure values for the anisotropic  and inhomogeneous diffusion induced by the Piola transformation only (top) and by the stress-assisted diffusion (centre). The right-bottom panel shows the anisotropy of the second Piola-Kirchhoff tensor. Ellipsoids with diverse sizes and shape ratios indicate a
  high anisotropic degree. The colour-bars indicate the pressure on the tensorial glyphs, whereas the field plotted in the back is the $L^2-$norm
  of each diffusion tensor.\label{fig:5}}
\end{figure}

The analysis is further quantified by rendering the modified diffusion
tensors in terms of their \emph{tensorial glyph} representation (see e.g.~\cite{tensorglyphs}).
These plots {(see Figure~\ref{fig:5})} are generated by
solving local eigenvalue problems using the tensor at hand and
displaying an ellipsoid whose shape and size determine the magnitude
of eigenvalues and direction of eigenvectors and indicating the local
patterns of anisotropy  and inhomogeneity. For reference, we also depict the $L^2-$norm of the 
resulting diffusion tensor in each case, that is 
$D_0 C_{IJ}^{-1}$, $D_{IJ}$, and $S_{IJ}$, respectively. 
As expected, constant diffusion, Figure~\ref{fig:5a}, and stress-assisted diffusion, Figure~\ref{fig:5b}, lead to different representations of the eigenproblem. 
Even if we  only account for the uniaxial case (represented in Figure~\ref{fig:5c}), 
substantial differences are expected when multiaxial stress patterns are considered.

According to \cite{panfilov:2005}, where stretch-activated currents are present, we observe a \emph{small} spiral drift velocity of about 1.4\% variation with respect to pulse wave velocity. This means that several spiral rotations are required to reach the boundary of the domain such to detect the exit of the excitation wave from the domain (auto-defibrillation). 
However such small differences can lead to substantially 
different scenarios at long times, as observed in strongly nonlinear dynamical systems
~\cite{bueno-orovio:2016,gizzi:2016}.  
This behaviour is in line with the expected vulnerability to atrial flutter and atrial fibrillation during atrial dilatation~\cite{ravelli:2005,mase:2008,jalife:2009}. 
At this stage we remark that our theoretical study does not imply a 
stretch-activated current contribution 
(still debated in the literature~\cite{quinn:2014})
but only relies on the effect induced by the generalised diffusivity tensor.

\begin{figure}[t!]
 \begin{center}
 \subfigure[]{\includegraphics[width=0.325\textwidth]{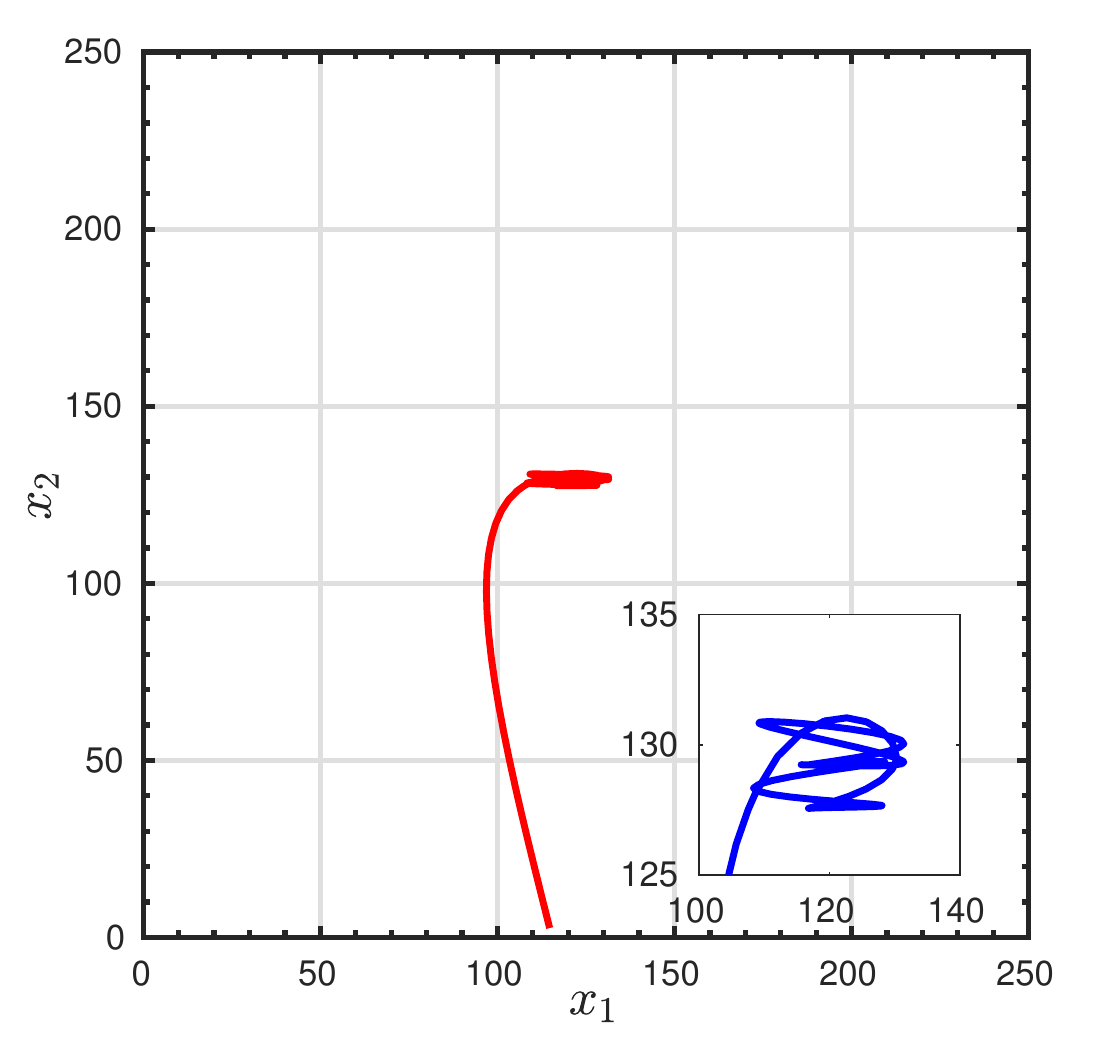}}
\subfigure[]{\includegraphics[width=0.325\textwidth]{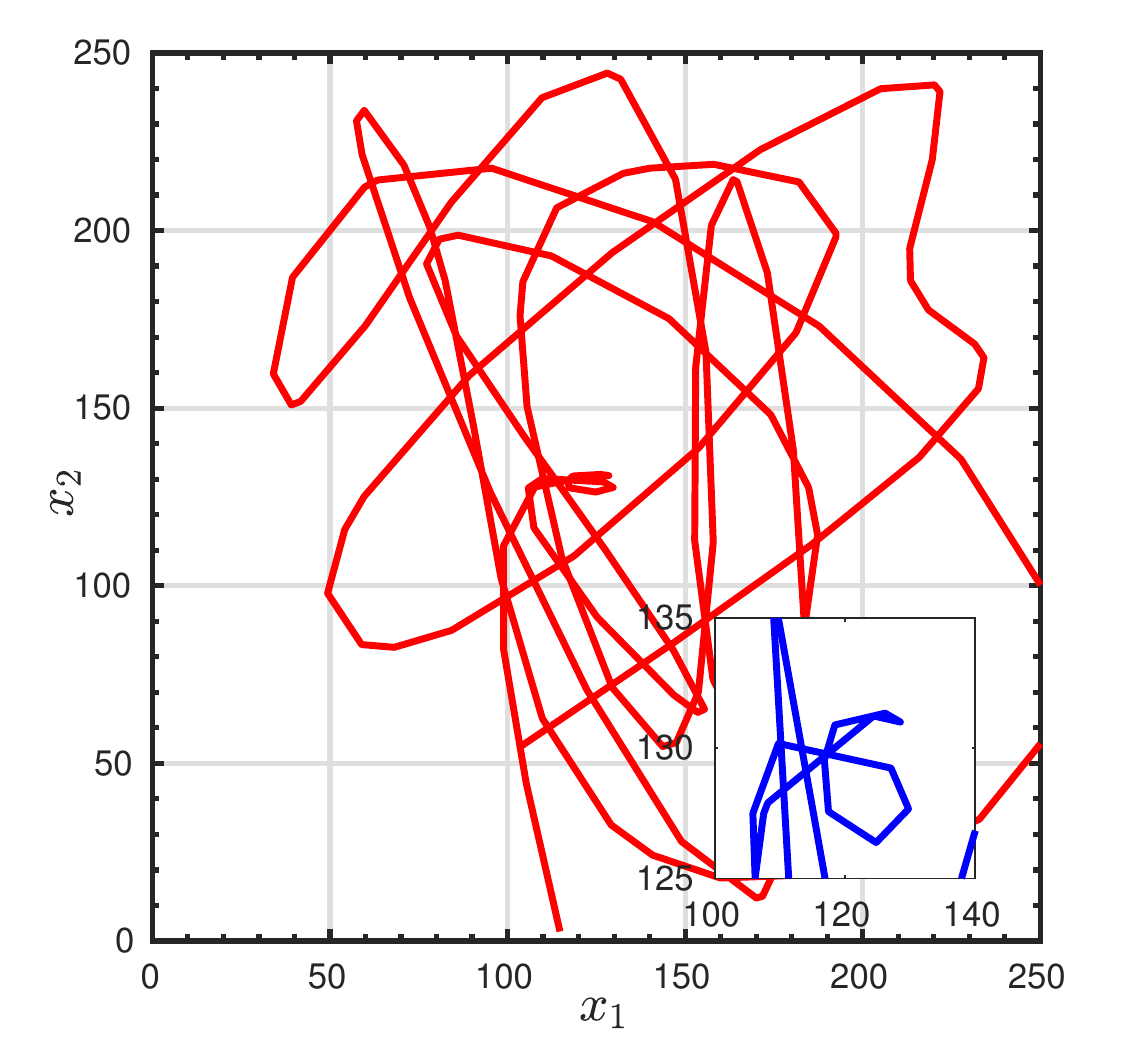}}
\subfigure[]{\includegraphics[width=0.325\textwidth]{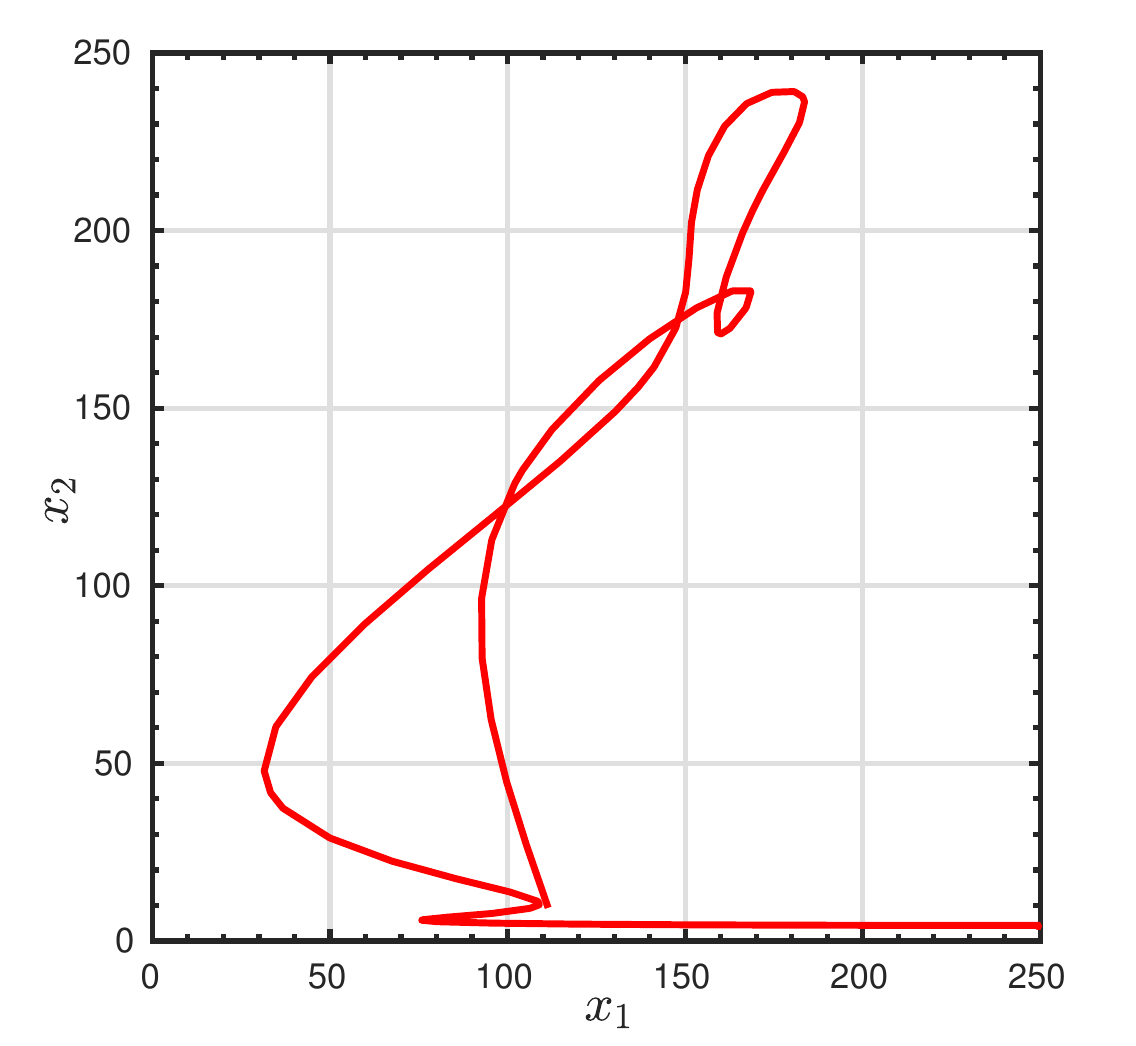}}
\end{center}
\vspace{-3mm}
 \caption{Enhanced effects due to mechanical pacing. Trajectories of the spiral tip depending on three different stress-assisted diffusion 
configurations.  $D_1=-0.015,D_2=0.015$ 
  (a), $D_1=0.275,D_2=-0.275$ (b), and  $D_1=-0.5,D_2=0.5$ (c). 
  The inset plots indicate 
  a zoom of each trajectory near the meandring location at larger times.\label{fig:tips}}
\end{figure}

% ****************************************************************************************
\paragraph{Enhanced effects due to mechanical pacing.}
Finally we carry out a set of simulations of the fully coupled 
stress-assisted diffusion model, concentrating in the case where a periodic and uniaxial 
mechanical loading is applied on the right end of the square slab $\Omega=(0,250)^2$. 
Again, the material and electrical parameters are taken from Table~\ref{table}. The 
boundary conditions for the nonlinear elasticity problem correspond to zero normal 
stresses on the horizontal boundaries, a clamped left boundary, and on the right we 
prescribe the displacement 
$$\bu =\left[50\sin^2\left(\frac{\pi}{50} [t-t^*]\right), \,0\right]^T,$$ 
where $t^*=100$ indicates the onset time for the mechanical loading. We use a S1-S2 protocol 
to initiate spiral wave dynamics, and focus our attention on the  long term 
behaviour of the patterns produced according to three different levels of stress-assisted diffusion.  
In Figure~\ref{fig:tips} we portray the trajectory of the spiral tip (also showing a zoom on the 
relevant region) over the time interval $t\in [400,1000]$. The location of the spiral tip 
is recorded by computing a piecewise constant field defined as the module 
of the gradient of the recovery variable, and then obtaining its global 
argmax over the spatial domain. Our results indicate a marked progressive drifting of the 
spiral tip, which increases as the relevant parameters depart from the stable values.

 % ****************************************************************************************
\section{Discussion}
In this work we propose a generalised MEF formulation for active deformable media 
based on the concept of stress-assisted diffusion. The leading mechanism of the modification of 
 wave propagation corresponds to the induction of anisotropy in an originally isotropic medium. In this study, we simulated a single spiral wave that drifts or degenerates according to the amount of stress-assisted coupling.  The proposed physics, experimentally and theoretically based on physical grounds, enriches the phenomenological description of electro-mechanical RD systems in the context of cardiac dynamics and excitable deformable media in general.
The numerical results confirm that the generalised
formulation induces anisotropy in the wave propagation starting from an isotropic medium. Secondly,  
the RD  dynamics are modified, as clearly evidenced  by the change in spiral meandering (i.e. drift of the spiral tip~\cite{nash:2004,panfilov:2005}). In addition, the generalised model leads to inhomogeneous and unpredicted excitation waves, especially pronounced 
for extreme values of the coupling coefficients as observed in long run tests and under mechanical pacing. 

To close this section we collect a few possible extensions of this work. 
The connection between spatio-temporal variation of the tissue anisotropy and the identification of model parameters  in general 
cases remains unclear, but it is by no means a trivial task. One would definitely require to analyse deformation patterns interacting with a number of 
intrinsic properties (not considered here) as domain curvature, a priori anisotropy, pre-stretch, structural heterogeneities, 
and many others.
\\
Also, we would be interested in analysing important microscopic effects related to gap junction proteins density at cell-cell interfaces. These properties have an important effect on the macroscopic diffusivity and its relation with the conduction velocity~\cite{thomas:2003}, especially under pathological remodelling conditions. Other envisaged modifications include 
different mechanical constraints~\cite{mccain:2012},  and the incorporation of 
beat-to-beat adaptation (see the recent review \cite{quinn:2016}).
\\
An important complement to our generalised MEF formulation would be the characterisation of admissibility ranges 
for an elementary mechanical setup, together with the experimental validation of the proposed 
model. The latter could be carried out by fine-tuning stress-assisted diffusion parameters via optical mapping 
of cell cultures,  myocardial tissue slices, and whole heart geometries (with particular attention to atrial tissue). \\
Several other applications are foreseen, as stretch-activated currents and 
MEF in cell-cell coupling phenomena. However a 
deeper understanding of the interacting regimes (also including model parameters  
and biophysical descriptions) is still required before engaging in the study of these mechanisms 
using homogenisation theory.

\paragraph{Acknowledgements.}
  This work has been partially supported by the Italian National Group of Mathematical Physics GNFM-INdAM, 
  by the International Center for Relativistic Astrophysics Network ICRANet, by the 
  London Mathematical Society through their Grant Scheme 4, and by the EPSRC through the Research Grant EP/R00207X/1. 

% ****************************************************************************************
\small 
\bibliographystyle{plain}
\bibliography{CFGR_newDraft.bib}

\providecommand{\noopsort}[1]{}\providecommand{\singleletter}[1]{#1}%
\begin{thebibliography}{10}

\bibitem{aifantis:1980}
E.~C. Aifantis.
\newblock On the problem of diffusion in solids.
\newblock {\em Acta Mechanica}, 37:265--296, 1980.

\bibitem{aliev:1996}
R.~R. Aliev and A.~V. Panfilov.
\newblock A simple two-variable model of cardiac excitation.
\newblock {\em Chaos Solitons Fractals}, 7:293, 1996.

\bibitem{bao:2004}
L.~Bao, F.~Sachs, and G.~Dahl.
\newblock Connexins are mechanosensitive.
\newblock {\em American Journal of Physiology - Cell Physiology},
  287(5):C1389--C1395, 2004.

\bibitem{Brezzi1985}
F.~Brezzi, J.~Douglas, and L.~D. Marini.
\newblock Two families of mixed finite elements for second order elliptic
  problems.
\newblock {\em Numerische Mathematik}, 47(2):217--235, 1985.

\bibitem{bueno-orovio:2016}
A.~Bueno-Orovio, I.~Teh, J.~E. Schneider, K.~Burrage, and V.~Grau.
\newblock Anomalous diffusion in cardiac tissue as an index of myocardial
  microstructure.
\newblock {\em IEEE Transactions Medical Imaging}, 35:2200--2207, 2016.

\bibitem{fenton:2017}
D.~D. Chen, R.~A. Gray, I.~Uzelac, C.~Herndon, and F.~H. Fenton.
\newblock Mechanism for amplitude alternans in electrocardiograms and the
  initiation of spatiotemporal chaos.
\newblock {\em Physical Review Letters}, page In Press., 2017.

\bibitem{chene:2008}
J.~Ch{\^e}ne.
\newblock {\em Environment-Induced Cracking of Materials}, chapter
  Strain-assisted transport of hydrogen and related effects on the
  intergranular stress corrosion cracking of alloy 600.
\newblock Elsevier Ltd, 2008.

\bibitem{cherian:2005}
P.~P. Cherian, A.~J. Siller-Jackson, S.~Gu, X.~Wang, L.~F. Bonewald,
  E.~Sprague, and J.~X. Jiang.
\newblock Mechanical strain opens connexin 43 hemichannels in osteocytes: A
  novel mechanism for the release of prostaglandin.
\newblock {\em Molecular Biology of the Cell}, 16(7):3100--3106, 2005.

\bibitem{cherubini:2008}
C.~Cherubini, S.~Filippi, P.~Nardinocchi, and L.~Teresi.
\newblock An electromechanical model of cardiac tissue: Constitutive issues and
  electrophysiological effects.
\newblock {\em Progress in Biophysics and Molecular Biology}, 97:562--573,
  2008.

\bibitem{cohen:1989}
D.~S. Cohen.
\newblock Sharp fronts due to diffusion and stress at the glass transition in
  polymers.
\newblock {\em Journal of Polymer Science: Part B: Polymer Physics},
  27:1731--1747, 1989.

\bibitem{das:2014}
N.~P. Das, D.~Mahanta, and S.~Dutta.
\newblock Unpinning of scroll waves under the influence of a thermal gradient.
\newblock {\em Physical Review E}, 90:022916, 2014.

\bibitem{mcculloch:2015}
B.~L. De~Oliveira, E.~R. Pfeiffer, S.~Joakim, S.~T. Wall, and A.~D. McCulloch.
\newblock Increased cell membrane capacitance is the dominant mechanism of
  stretch-dependent conduction slowing in the rabbit heart: a computational
  study.
\newblock {\em Cellular and Molecular Bioengineering}, 8:237--246, 2015.

\bibitem{tensorglyphs}
D.~B. Ennis, G.~Kindlman, I.~Rodriguez, P.~A. Helm, and E.~R. McVeigh.
\newblock Visualization of tensor fields using superquadric glyphs.
\newblock {\em Magnetic Resonance in Medicine}, 53(1):169--176, 2005.

\bibitem{franz:1989}
M.~R. Franz, D.~Burkhoff, D.~T. Yue, and K.~Sagawa.
\newblock Mechanically induced action potential changes and arrhythmia in
  isolated and in situ canine hearts.
\newblock {\em Cardiovascular Research}, 23:213--223, 1989.

\bibitem{gizzi:2015}
A.~Gizzi, C.~Cherubini, S.~Filippi, and A.~Pandolfi.
\newblock Theoretical and numerical modeling of nonlinear electromechanics with
  applications to biological active media.
\newblock {\em Communications in Computational Physics}, 17:93--126, 2015.

\bibitem{suo:2009}
W.~Hong, Z.~Liu, and Z.~Suo.
\newblock Inhomogeneous swelling of a gel in equilibrium with a solvent and
  mechanical load.
\newblock {\em International Journal of Solids and Structures}, 49:3282--3289,
  2009.

\bibitem{gizzi:2016}
D.~E. Hurtado, S.~Castro, and A.~Gizzi.
\newblock Computational modeling of non-linear diffusion in cardiac
  electrophysiology: A novel porous-medium approach.
\newblock {\em Computer Methods in Applied Mechanics and Engineering},
  300:70--83, 2016.

\bibitem{jackson:2009}
A.~R. Jackson, F.~Travascio, and W.~Y. Hu.
\newblock Effect of mechanical loading on electrical conductivity in human
  intervertebral disc.
\newblock {\em Journal of Biomechanical Engineering}, 131:054505, 2009.

\bibitem{johnson:1996}
P.~A. Johnson and P.~N.~J. Rasolofosaon.
\newblock Nonlinear elasticity and stress-induced anisotropy in rock.
\newblock {\em Journal of Geophysical Research}, 101:3113--3124, 1996.

\bibitem{kaufmann:1967}
R.~Kaufmann and U.~Theophile.
\newblock Automatic-fordernde dehnungseffekte an purkinje-faden,
  pappillarmuskeln und vorhoftrabekeln von rhesus-affen.
\newblock {\em Pflugers Arch. Gesamte Physiol. Menschen Tiere}, 297:174--189,
  1967.

\bibitem{klepach:2014}
D.~Klepach and T.~I. Zohdi.
\newblock Strain assisted diffusion: Modeling and simulation of
  deformation-dependent diffusion in composite media.
\newblock {\em Composites: Part B}, 56:413--423, 2014.

\bibitem{kohl:2001}
P.~Kohl and F.~Sachs.
\newblock Mechanoelectric feedback in cardiac cells.
\newblock {\em Phylosofical Transactions of the Royal Society London A},
  359:1173--1185, 2001.

\bibitem{lab:1978}
M.~J. Lab.
\newblock Mechanically dependent changes in action potentials recorded from the
  intact frog ventricle.
\newblock {\em {Circulation Research}}, 42:519--528, 1978.

\bibitem{landau:1984}
L.~D. Landau, E.~M. Lifshitz, and L.~P. Pitaevskii.
\newblock {\em Electrodynamics of Continuous Media}, volume~8.
\newblock Butterworth-Heinemann, Oxford, 1984.

\bibitem{mase:2008}
M.~Mas\'e and F.~Ravelli.
\newblock A model for mechano-electrical feedback effects on atrial flutter
  interval variability.
\newblock {\em Bullettin of Mathematical Biology}, 70:1326--1347, 2008.

\bibitem{mccain:2012}
M.~L. McCain, H.~Lee, Y.~Aratyn-Schaus, A.~G. Kl{\'e}ber, and K.~K. Parker.
\newblock Cooperative coupling of cell-matrix and cell--cell adhesions in
  cardiac muscle.
\newblock {\em Proceedings of the National Academy of Sciences},
  109(25):9881--9886, 2012.

\bibitem{miehe:2014}
C.~Miehe, S.~Mauthe, and H.~Ulmer.
\newblock Formulation and numerical exploitation of mixed variational
  principles for coupled problems of cahn--hilliard-type and standard diffusion
  in elastic solids.
\newblock {\em International Journal for Numerical Methods in Engineering},
  99:737--762, 2014.

\bibitem{miller:2002}
B.~S. Miller-Chou and J.~L. Koenig.
\newblock A review of polymer dissolution.
\newblock {\em Progress in Polymer Science}, 28:1223--1270, 2002.

\bibitem{mills:2011}
R.~W. Mills, A.~T. Wright, S.~M. Narayan, and A.~D. McCulloch.
\newblock {\em The effects of wall stretch on ventricular conduction and
  refractoriness in the whole heart}, chapter~25.
\newblock Oxford University Press, 2011.

\bibitem{nash:2004}
M.~P. Nash and A.~V. Panfilov.
\newblock Electromechanical model of excitable tissue to study reentrant
  cardiac arrhythmias.
\newblock {\em Progress in Biophysics and Molecular Biology}, 85:501--522,
  2004.

\bibitem{nava:2016}
M.~M. Nava, R.~Fedele, and M.~T. Raimondi.
\newblock Computational prediction of strain-dependent diffusion of
  transcription factors through the cell nucleus.
\newblock {\em Biomechanics and Modeling in Mechanobiology}, 15:983--993, 2016.

\bibitem{nobile:2012}
F.~Nobile, R.~Ruiz-Baier, and A.~Quarteroni.
\newblock An active strain electromechanical model for cardiac tissue.
\newblock {\em International Journal for Numerical Methods in Biomedical
  Engineering}, 28:52--71, 2012.

\bibitem{ogden}
R.~W. Ogden.
\newblock {\em Non-Linear Elastic Deformations}.
\newblock Dover Publications, 1997.

\bibitem{orourke:1999}
B.~O'Rourke, D.~A. Kass, G.~F. Tomaselli, S.~K\"a\"ab, R.~Tunin, and
  E.~Marb\'an.
\newblock Mechanisms of altered excitation-contraction coupling in canine
  tachycardia-induced heart failure, i experimental studies.
\newblock {\em {Circulation Research}}, 84:562--570, 1999.

\bibitem{panfilov:2005}
A.~V. Panfilov and R.~H. Keldermann.
\newblock Self-organized pacemakers in a coupled reaction-diffusion-mechanics
  system.
\newblock {\em Physical Review Letters}, 95:258104, 2005.

\bibitem{panfilov:2007}
A.~V. Panfilov, R.~H. Keldermann, and M.~P. Nash.
\newblock Drift and breakup of spiral waves in reaction-diffusion-mechanics
  systems.
\newblock {\em Proceedings of the National Academy of Sciences},
  104:7922--7926, 2007.

\bibitem{quinn:2014a}
T.~A. Quinn.
\newblock The importance of non-uniformities in mechano-electric coupling for
  ventricular arrhythmias.
\newblock {\em J Interv Card Electrophysiol}, 39:25--35, 2014.

\bibitem{quinn:2015}
T.~A. Quinn.
\newblock Cardiac mechano-electric coupling: a role in regulating normal
  function of the heart?
\newblock {\em Cardiovascular Research}, 108:1--3, 2015.

\bibitem{quinn:2013}
T.~A. Quinn and P.~Kohl.
\newblock Combining wet and dry research: experience with model development for
  cardiac mechano-electric structure-function studies.
\newblock {\em Cardiovascular Research}, 97:601--611, 2013.

\bibitem{quinn:2016}
T.~A. Quinn and P.~Kohl.
\newblock Rabbit models of cardiac mechano-electric and mechano-mechanical
  coupling.
\newblock {\em Progress in Biophysics and Molecular Biology}, 121(2):110 --
  122, 2016.

\bibitem{quinn:2014}
T.~A. Quinn, P.~Kohl, and U.~Ravens.
\newblock Cardiac mechano-electric coupling research: Fifty years of progress
  and scientific innovation.
\newblock {\em Progress in Biophysics and Molecular Biology}, 115:71--75, 2014.

\bibitem{ravelli:2003}
F.~Ravelli.
\newblock Mechano-electric feedback and atrial fibrillation.
\newblock {\em Progress in Biophysics and Molecular Biology}, 82:137--149,
  2003.

\bibitem{ravelli:2005}
F.~Ravelli, L.~Faes, L.~Sandrini, F.~Gaita, R.~Antolini, M.~Scaglione, and
  G.~Nollo.
\newblock Wave similarity mapping shows the spatiotemporal distribution of
  fibrillatory wave complexity in the human right atrium during paroxysmal and
  chronic atrial fibrillation.
\newblock {\em Journal of Cardiovascular Electrophysiology}, 16:1071--1076,
  2005.

\bibitem{Raviart1977}
P.~A. Raviart and J.~M. Thomas.
\newblock {\em A mixed finite element method for 2-nd order elliptic problems},
  pages 292--315.
\newblock Springer Berlin Heidelberg, 1977.

\bibitem{ruiz:2015}
R.~Ruiz-Baier.
\newblock Primal-mixed formulations for reaction-diffusion systems on deforming
  domains.
\newblock {\em Journal of Computational Physics}, 299:320--338, 2015.

\bibitem{salameh:2013}
A.~Salamhe and S.~Dhein.
\newblock Effects of mechanical forces and stretch on intercellular gap
  junction coupling.
\newblock {\em Biochimica et Biophysica Acta (BBA) - Biomembranes}, 1828(1):147
  -- 156, 2013.

\bibitem{showalter}
R.E. Showalter.
\newblock Degenerate parabolic initial-boundary value problems.
\newblock {\em Journal of Differential Equations}, 31(3):296--312, 1979.

\bibitem{spencer:1980}
A.~J.~M. Spencer.
\newblock {\em Continuum Mechanics}.
\newblock Longman Group Ltd, London, 1989.

\bibitem{thomas:2003}
Stuart~P. Thomas, Jan~P. Kucera, Lilly Bircher-Lehmann, Yoram Rudy, Jeffrey~E.
  Saffitz, and Andr{\'e}~G. Kl{\'e}ber.
\newblock Impulse propagation in synthetic strands of neonatal cardiac myocytes
  with genetically reduced levels of connexin 43.
\newblock {\em Circulation Research}, 92(11):1209--1216, 2003.

\bibitem{weiss:2009}
D.~X. Tran, D.~Sato, A.~Yochelis, J.~N. Weiss, A.~Garfinkel, and Z.~Qu.
\newblock Bifurcation and chaos in a model of cardiac early
  afterdepolarizations.
\newblock {\em Physical Review Letters}, 102:258103, 2009.

\bibitem{trayanova:2011}
N.~A. Trayanova, J.~Costantino, and V.~Gurev.
\newblock Electromechanical models of the ventricles.
\newblock {\em Am. J. Physiol. Heart Circ. Physiol.}, 301:H279--H286, 2011.

\bibitem{suo:1997}
W.~Wang and Z.~Suo.
\newblock Shape change of a pore in a stressed solid via surface diffusion
  motivated by surface and elastic energy variation.
\newblock {\em Journal of the Mechanics and Physics of Solids}, 45:709--729,
  1997.

\bibitem{weitsman:1987}
Y.~Weitsman.
\newblock Stress assisted diffusion in elastic and viscoelastic materials.
\newblock {\em Journal of the Mechanics and Physics of Solids}, 35:73--93,
  1987.

\bibitem{jalife:2009}
M.~Yamazaki, L.~M. Vaquero, L.~Hou, K.~Campbell, S.~Zlochiver, M.~Klos,
  S.~Mironov, O.~Berenfeld, H.~Honjo, I.~Kodama, J.~Jalife, and J.~Kalifa.
\newblock Mechanisms of stretch induced atrial fibrillation in the presence and
  the absence of adreno-cholinergic stimulation: Interplay between rotors and
  focal discharges.
\newblock {\em Heart Rhythm}, 6:1009--1017, 2009.

\bibitem{yuan:2009}
T.~Y. Yuan, A.~R. Jackson, C.~Y. Huang, and W.~Y. Gu.
\newblock Strain-dependent oxygen diffusivity in bovine annulus fibrosus.
\newblock {\em Journal of Biomechanical Engineering}, 131:074503, 2009.

\bibitem{zheng:2015}
Y.~T. Zheng, F.~Z. Xuan, and Z.~Wang.
\newblock A dominant role of stress-dependent oxide structure on diffusion flux
  in the strain-reaction engineering.
\newblock {\em Chemical Physics Letters}, 626:25--28, 2015.

\end{thebibliography}

\end{document}